\documentclass[a4paper,twocolumn,11pt,accepted=2021-08-31]{quantumarticle}
\pdfoutput=1

\usepackage{graphicx,bm,amsmath,eufrak,amssymb,url}
\usepackage[usenames,dvipsnames]{color}
\usepackage[colorlinks,bookmarks=false,citecolor=NavyBlue,linkcolor=Red,urlcolor=blue]{hyperref}
\usepackage[bbgreekl]{mathbbol}
\usepackage{xcolor}
\usepackage{hyperref}
\usepackage{soul}

\usepackage{subcaption}

\usepackage[T1]{fontenc}
\usepackage[latin9]{inputenc}
\usepackage[english]{babel} 
\usepackage{verbatim}
\usepackage{graphicx}

 \usepackage{grffile}
\usepackage[numbers,sort&compress]{natbib}

\usepackage{bbold}

\newcommand{\be}{\begin{equation}}
\newcommand{\ee}{\end{equation}}
\newcommand{\bea}{\begin{eqnarray}}
\newcommand{\eea}{\end{eqnarray}}

\definecolor{smoothred}{HTML}{C5232F}
\definecolor{mygreen}{rgb}{0,0.5,0}
\definecolor{myblue}{rgb}{0,0,0.75}
\definecolor{mymagenta}{cmyk}{0,1,0,0.12}

\def\doi{http://dx.doi.org/}

\begin{document}

\title{Hilbert curve vs Hilbert space: 
exploiting fractal 2D covering to increase tensor network efficiency
}

\author{Giovanni~Cataldi}
\affiliation{Dipartimento di Fisica e Astronomia ``G. Galilei'', Universit\`a di Padova, I-35131 Padova, Italy.}
\affiliation{Padua Quantum Technologies Research Center, Universit\`a degli Studi di Padova.}
\affiliation{Istituto Nazionale di Fisica Nucleare (INFN), Sezione di Padova, I-35131 Padova, Italy.}  

\author{Ashkan~Abedi}
\affiliation{Scuola Normale Superiore, I-56127 Pisa, Italy.}

\author{Giuseppe~Magnifico}
\affiliation{Dipartimento di Fisica e Astronomia ``G. Galilei'', Universit\`a di Padova, I-35131 Padova, Italy.}
\affiliation{Padua Quantum Technologies Research Center, Universit\`a degli Studi di Padova.}
\affiliation{Istituto Nazionale di Fisica Nucleare (INFN), Sezione di Padova, I-35131 Padova, Italy.}  

\author{Simone~Notarnicola}
\affiliation{Dipartimento di Fisica e Astronomia ``G. Galilei'', Universit\`a di Padova, I-35131 Padova, Italy.}
\affiliation{Padua Quantum Technologies Research Center, Universit\`a degli Studi di Padova.}
\affiliation{Istituto Nazionale di Fisica Nucleare (INFN), Sezione di Padova, I-35131 Padova, Italy.}  

\author{Nicola~Dalla~Pozza}
\affiliation{Scuola Normale Superiore, I-56127 Pisa, Italy.}

\author{Vittorio~Giovannetti}
\affiliation{NEST, Scuola Normale Superiore and Istituto Nanoscienze-CNR, I-56127 Pisa, Italy.}

\author{Simone~Montangero}
\affiliation{Dipartimento di Fisica e Astronomia ``G. Galilei'', Universit\`a di Padova, I-35131 Padova, Italy.}   
\affiliation{Padua Quantum Technologies Research Center, Universit\`a degli Studi di Padova.}
\affiliation{Istituto Nazionale di Fisica Nucleare (INFN), Sezione di Padova, I-35131 Padova, Italy.}


\begin{abstract}
We present a novel mapping for studying 2D many-body quantum systems by solving an effective, one-dimensional long-range model in place of the original two-dimensional short-range one. In particular, we address the problem of choosing an efficient mapping from the 2D lattice to a 1D chain that optimally preserves the
locality of interactions within the TN structure. By using Matrix Product States (MPS)  and Tree Tensor Network (TTN) algorithms, we compute the ground state of the 2D quantum Ising model in transverse field with lattice 
size up to $64\times64$, comparing the results obtained from different mappings based on two space-filling curves, the snake curve and the Hilbert curve. We show that the locality-preserving properties of the Hilbert curve leads to a clear improvement of numerical precision, especially for large sizes, and turns out to provide the best performances for the simulation of 2D lattice systems via 1D TN structures.
\end{abstract}

\maketitle

Unveiling the static and dynamical properties of strongly correlated quantum many-body systems remains one of the fundamental goals of the current research in condensed-matter physics.  Quantum systems with a large number of interacting microscopic constituents can generate collective states that have no classical counterpart and are extremely interesting both from a theoretical point of view as well as for possible experimental applications \cite{sachdev_2011, 102307jctt19cc2gc, girvin_yang_2019}. 
Paradigmatic examples include quantum Hall states \cite{RevModPhys.89.025005}, superconductors \cite{Rev_Superconductors}, topological insulators  \cite{Rachel_2018} and quantum spin liquids \cite{Savary_2016}. Such collective quantum phenomena play a central role not only in condensed-matter but also in high-energy physics, quantum chemistry, atomic physics as well as in quantum technologies where it is of paramount importance to understand and control the quantum states of many-body systems for 
quantum simulation and computation \cite{Review_QuantumTech, alexeev2020quantum, RevModPhys.86.153}. 

\begin{figure}[t!]
    \centering
     \begin{subfigure}[b]{0.49\columnwidth}
      \captionsetup{position=top, singlelinecheck=off,justification=raggedright}
     \caption{}
         \centering
         \includegraphics[width=\columnwidth]{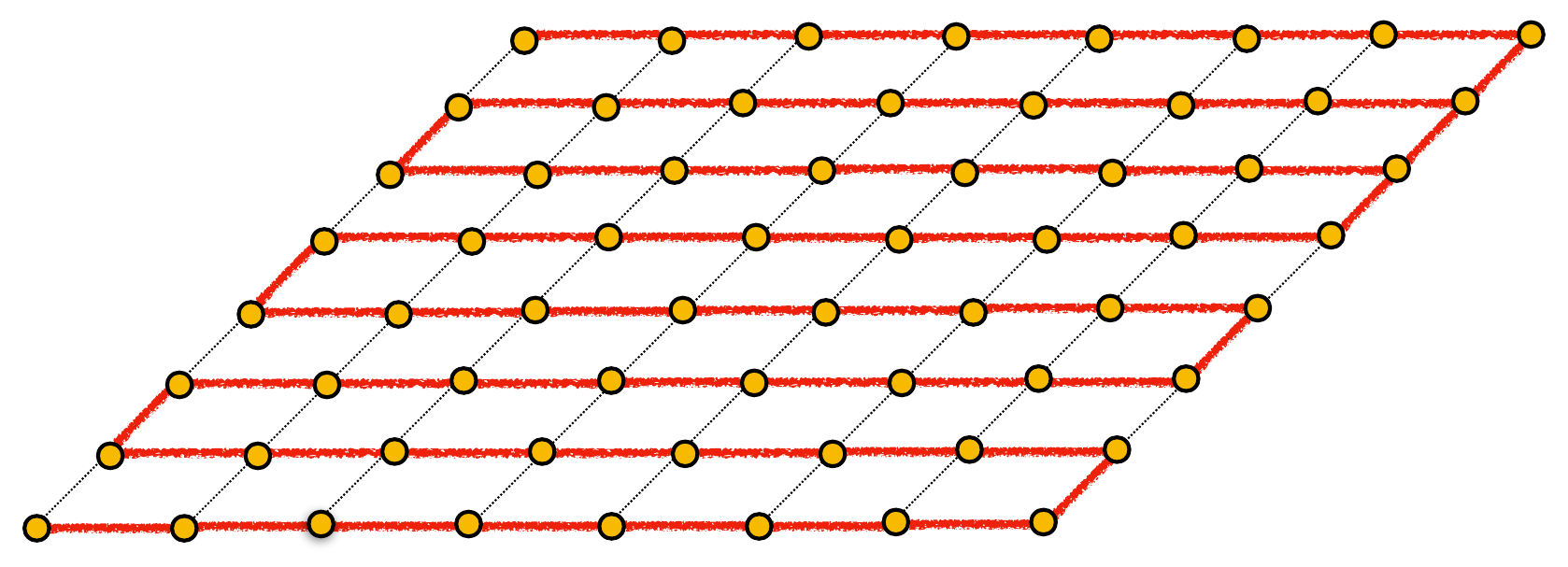}
         \label{fig:Snake_Curve_pictorial}
     \end{subfigure}
     \hfill
     \begin{subfigure}[b]{0.49\columnwidth}
      \captionsetup{position=top, singlelinecheck=off,justification=raggedright}
     \caption{}
         \centering
         \includegraphics[width=\columnwidth]{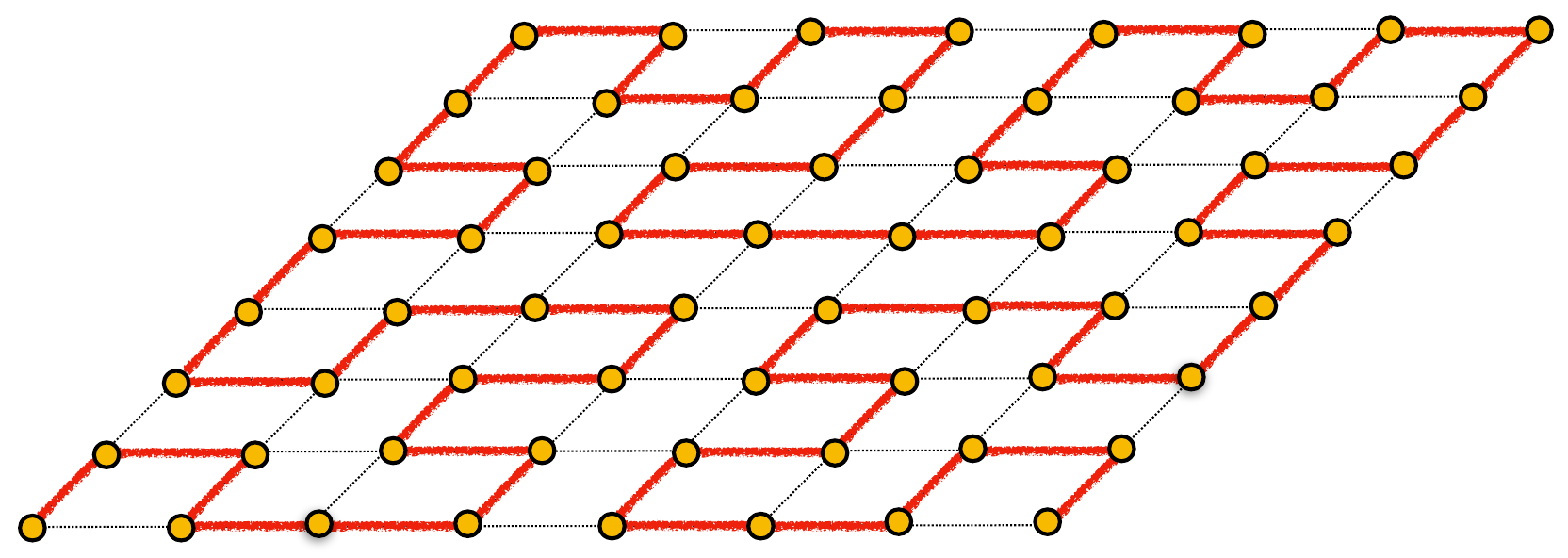}
         \label{fig:Hilbert_Curve_pictorial}
     \end{subfigure}
          \hfill
     \begin{subfigure}[b]{0.47\columnwidth}
      \captionsetup{position=top,singlelinecheck=off,justification=raggedright}
      \caption{}
         \centering
          \includegraphics[width=\columnwidth]{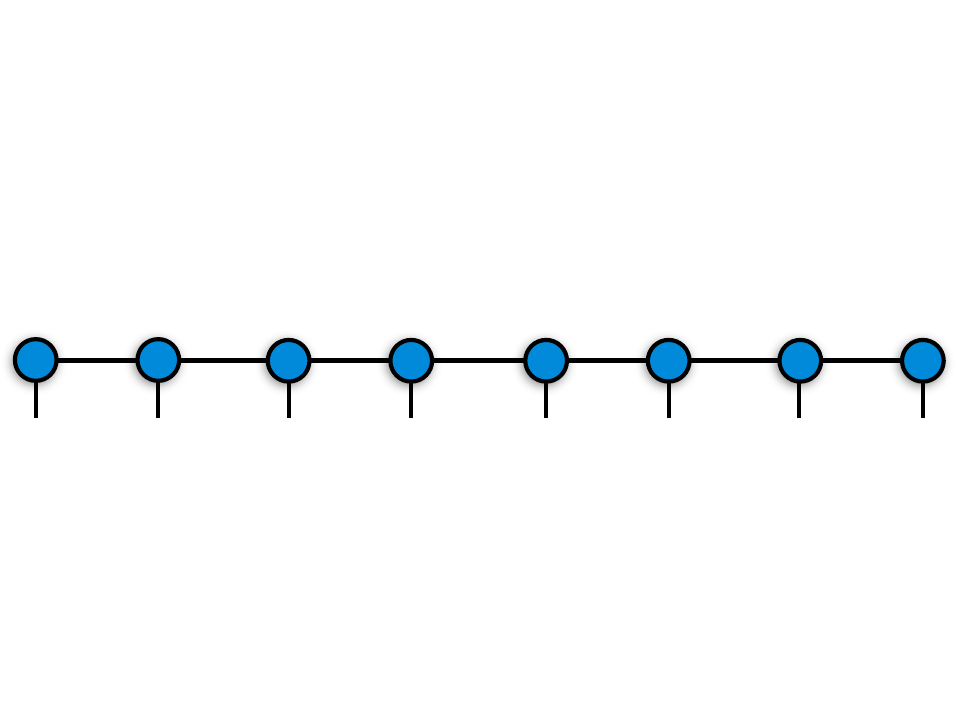}
         \label{fig:1D_MPS}
     \end{subfigure}
     \hfill
          \begin{subfigure}[b]{0.48\columnwidth}
      \captionsetup{position=top,singlelinecheck=off,justification=raggedright}
      \caption{}
         \centering
          \includegraphics[width=\columnwidth]{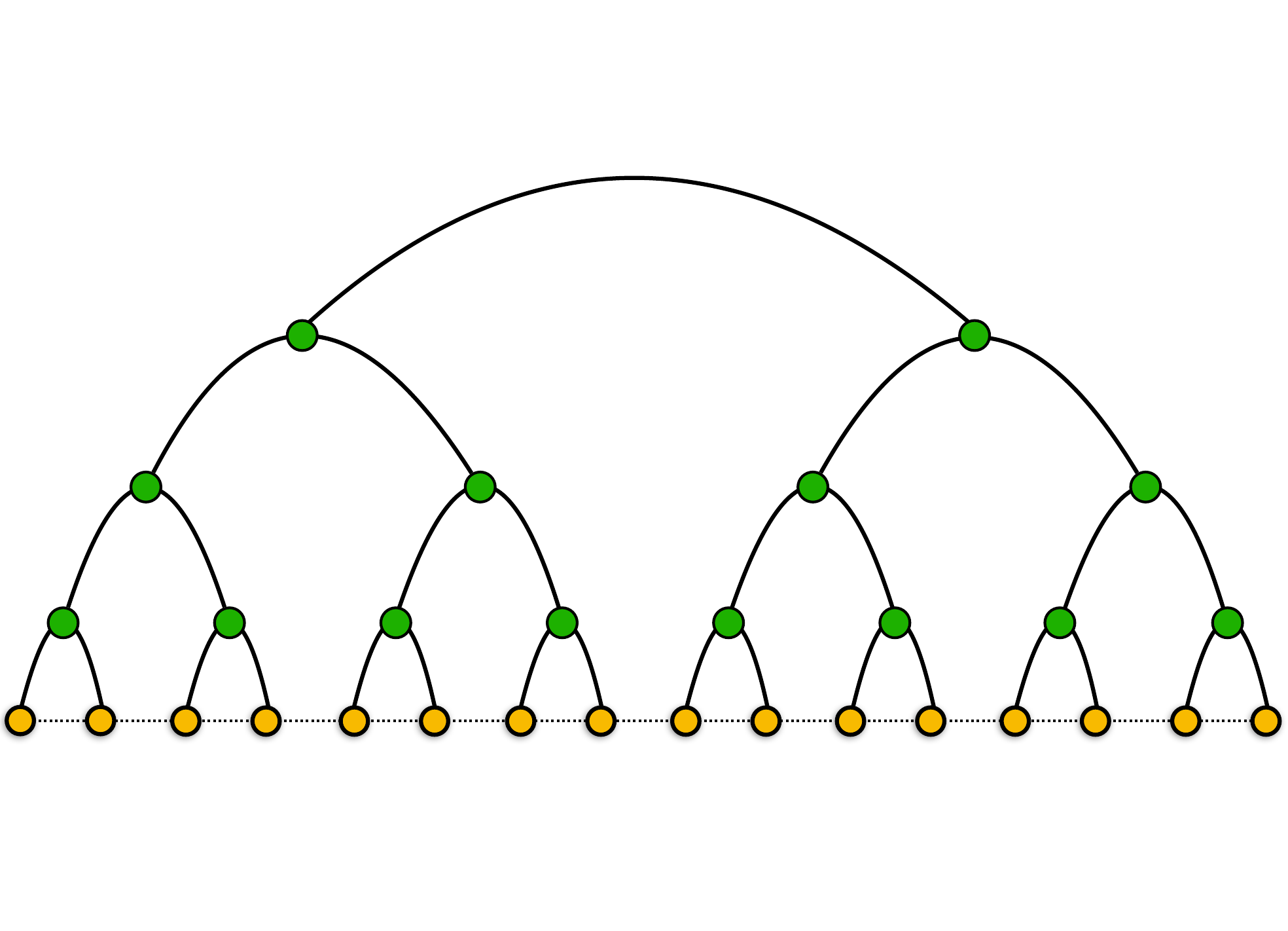}
         \label{fig:1D_TTN}
     \end{subfigure}
      \captionsetup{justification=centerlast}
\caption{Pictorial representation of the Snake (a) and the Hilbert curve (b) for a 2D lattice with linear size $n=8$. (c) Standard MPS representation for 1D chain. (d) Binary TTN Ansazt for 1D lattice. Yellow circles represent the physical sites of the lattice, whereas the green circles indicate the tensors that make up the TTN. }
\end{figure}
Beyond analytical approaches and exact solutions, that are typically limited to a small class of Hamiltonians and {\color{black} microscopic} interactions, numerical methods play a key role in characterizing the low-energy properties 
of strongly correlated quantum many-body systems, characterizing their phase diagrams and the respective phase transitions \cite{scalapino1998numerical, 2008_Computational_Physics_Book}.
Despite their impressive success and accuracy in many scenarios, the complete understanding of several phenomena, e.g. the low-energy properties of Hubbard-like Hamiltonians for interacting fermions or the behavior of spin systems with frustrating interactions, is still debated, especially in two or three spatial dimensions \cite{doi10.11428676, PhysRevX.5.041041, PhysRevX.10.031016, PhysRevLett.111.137204, Stapmanns_2018, Maksimov_2019, czarnik2020tensor, wietek2020stripes},  where current and near-future 
experimental setups require benchmarking tools {\color{black} capable of simulating} large quantum many-body systems \cite{Scholl2020,Ebadi2020,Blatt2012,Henriet2020}.

In recent years, these challenging problems have inspired the development 
of novel numerical algorithms based on Tensor Networks (TN), which do not 
suffer from the well-known sign problem affecting Quantum Monte Carlo (QMC) methods \cite{doi10.108014789940801912366, PhysRevB.94.165116, Montangero_book, 2019NatRP1538O}. 
These techniques exploit the area-law entanglement bounds satisfied by a large class of physical systems  and allow an efficient 
representation of the low-energy sector by retaining in memory only the relevant states that contribute to the equilibrium properties and the real-time dynamics \cite{Eisert_2010}. In particular, this is carried out by compressing the exponentially large wavefunctions into a network of tensors interconnected through auxiliary indices with bond dimension $m$.
The main {\it Anz\"atze} for the representation of quantum many-body states based on TNs include Matrix Product States (MPS) for 1D systems \cite{MPS_1, MPS_2, MPS_3}, Projected Entangled Pair States (PEPS) \cite{PhysRevLett.96.220601, 2004cond.mat.7066V, tepaske2020threedimensional}, Tree Tensor Networks (TTN) \cite{PhysRevA.74.022320, PhysRevB.90.125154, PhysRevA.81.062335, TN_Anthology} and Multiscale Entanglement Renormalization Ansatz (MERA) \cite{PhysRevLett.99.220405, PhysRevLett.102.180406} which can be defined in any dimension. However, while for one-dimensional systems MPS are the established TN ansatz, the development of efficient TN algorithms for higher-dimensional systems is still ongoing \cite{PhysRevB.92.035142, PhysRevB.94.155123, PhysRevB.98.235148, PhysRevB.96.045128, PhysRevX.8.031031, PhysRevLett.122.070502, felser2020efficient, magnifico2020lattice}.

In this framework, a possible and frequently used approach lies in the application of consolidated one-dimensional numerical methods, {\color{black} such as the Density Matrix Renormalization Group (DMRG) algorithm \cite{PhysRevLett.69.2863, RevModPhys.77.259}}, to the study of higher-dimensional systems, by suitably mapping the high-dimensional problem onto an equivalent one-dimensional one. {\color{black} Indeed}, DMRG algorithm, despite being defined for 1D systems, has long been used for analysing two-dimensional quantum many-body systems, providing important and reliable numerical results in a wide variety of condensed-matter problems \cite{2D_DMRG_review}. Two main strategies can be implemented: the first one consists in considering the $n \times n'$ 2D lattice  as an $n'$-legs ladder, then grouping together all the $n'$ 
 sites of each column in a single numerical site, and finally applying the one-dimensional algorithms to the resulting chain \cite{10.21468/SciPostPhys.6.3.028}. 
 The price to pay in this case is the exponential growth with $n'$ of the local basis  dimension, which usually limits the application of this strategy to quasi-two-dimensional systems, in which the number of ladder legs is small with respect to the longitudinal extension.
The second strategy, on which we focus in this work, lies in covering the 
two-dimensional lattice by a one-dimensional curve and then applying one-dimensional algorithms to the resulting effective chain. The drawback at this approach is that even if the original model contains only nearest-neighbor interactions, the resulting model on the effective one-dimensional 
chain shows long-range interactions, that have an influence on the numerical approach efficiency \cite{6114433}.

In detail, the latter strategy induces a specific one-dimensional site-ordering in the 2D lattice, generating a mapping between the 2D physical system and a 1D chain. This procedure can be easily defined for any mapping
\begin{equation}\label{eq:mapping}
\mathcal{M}:\,i\in [1,n]\times [1,n'] \rightarrow \mu \in [1,nn']
\end{equation}
from a $n\times n'$ lattice onto a chain with $n\,n'$ sites, through which it is possible to translate a two-dimensional model into a one-dimensional one. 
Finally, one can apply 1D TN algorithms, such as DMRG, taking into account the long-range interactions {{resulting from}} the mapping \cite{PhysRevB.49.9214, PhysRevB.64.104414, PhysRevB.100.075138, PhysRevB.101.085124,ali2021ordering}. 
{Long-range interactions eventually  require a larger bond dimension to properly describe the system. }It follows that the performance of the simulation  depends on the capability of the mapping to preserve the locality of the interactions from the original 2D Hamiltonian to the effective new 1D one.

In this paper, we quantify the importance of the choice of the site-ordering curve.
In particular, we consider a square $n\times n$ lattice and compare the precision we achieve by using  the standard snake curve (see Fig. \ref{fig:Snake_Curve_pictorial}) and the {\it Hilbert curve} (see Fig. \ref{fig:Hilbert_Curve_pictorial}) \cite{Hilbert_Curve1}: as we discuss in detail later, the peculiar features of the Hilbert curve turn out to be very useful for simulating systems of large sizes via TN methods, especially close to a quantum phase transition where a divergence of the correlation length  is expected. 

We present the numerical results for the 2D quantum Ising model in transverse field  obtained by using the  {{Hilbert curve}} and the snake curve \cite{2013LNP862S}. 
The simulations are performed by using MPSs (see Fig. \ref{fig:1D_MPS})   
and binary TTNs (see Fig. \ref{fig:1D_TTN}). The latter are particularly interesting for a number of reasons \cite{TN_Anthology}: (i) with their hierarchical structure, they offer a strong connectivity such that the distance between two lattice sites, in terms of links connecting them, scales only logarithmically within the network, whereas for MPS the distance is clearly linear. As a results, TTNs are particularly advantageous to describe systems with long-range interactions with both open (OBC) and periodic (PBC) boundary conditions; (ii) similarly to MPS, they provide an efficient basis for the DMRG algorithm, which represent the state-of-the-art technique for the numerical simulation of quantum many-body 
systems in 1D with both local and long-range interactions; (iii) they have a loop-free structure and, as a result, they show a low algorithmic complexity ($O(m^4)$), that is slightly larger than MPS ($O(m^3)$), but significantly lower than intrinsic 2D TN algorithms such as PEPS or MERA, that suffer from very high computational costs (at least $O(m^{10})$).

In order to test the performance obtained with the two mappings, we focus 
on the ground state properties of the Ising model at zero temperature: we 
numerically determine the ground state energy at different lattice sizes up to $n=64$ with OBC. We find that the simulations performed with the Hilbert curve mapping provide more accurate ground state energies, for both MPS and TTN algorithms. We also find that the Hilbert curve shows a faster convergence of the energy than the snake one as the bond dimension is increased. Finally, we remarkably observe a larger improvement close to the critical point, stressing the relevance of our analysis for the study of {{quantum phase transitions.}}
 Analogous results are also provided for the PBC case with TTN and are shown in the Appendix \ref{app:PBC_TTN}.

The article is organized as follows. In Sec. \ref{sec:Hilbert_Curve} we introduce the reader to the Hilbert and snake curves, briefly reviewing their space-filling and locality-preserving properties. In Sec. \ref{sec:quantum_ising_model} we describe the 2D quantum Ising model we are going to 
focus on, while Sec. \ref{sec:num_res} is devoted to the presentation and 
discussion of the numerical results we obtain, with the MPS and TTN algorithms. In Sec. \ref{sec:conclusions} we   draw our conclusions. Supplementary technical details are given in the Appendix.

\section{The Hilbert Curve }\label{sec:Hilbert_Curve}

\begin{figure}[t!]
        \centering
        \begin{subfigure}[b]{0.47\columnwidth}
            \centering
              \captionsetup{position=top,singlelinecheck=off,justification=raggedright}
              \caption{}
            \includegraphics[width=\columnwidth]{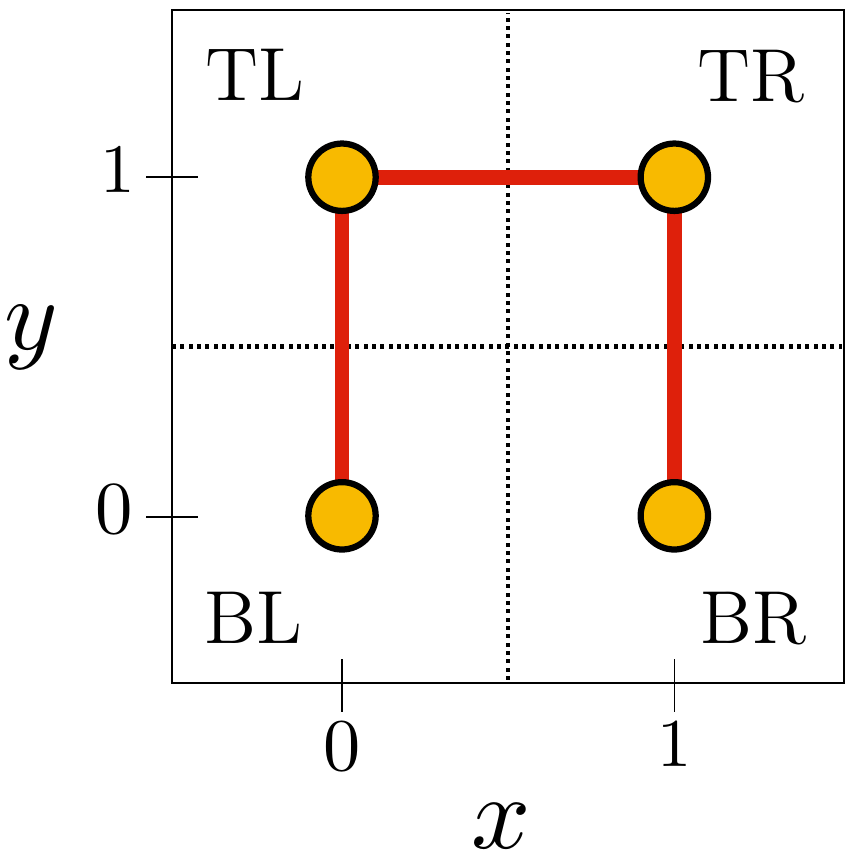}
            \label{fig:Hilbert_2}
        \end{subfigure}
        \hfill
        \begin{subfigure}[b]{0.47\columnwidth}  
            \centering 
             \captionsetup{position=top,singlelinecheck=off,justification=raggedright}
             \caption{}
            \includegraphics[width=\columnwidth]{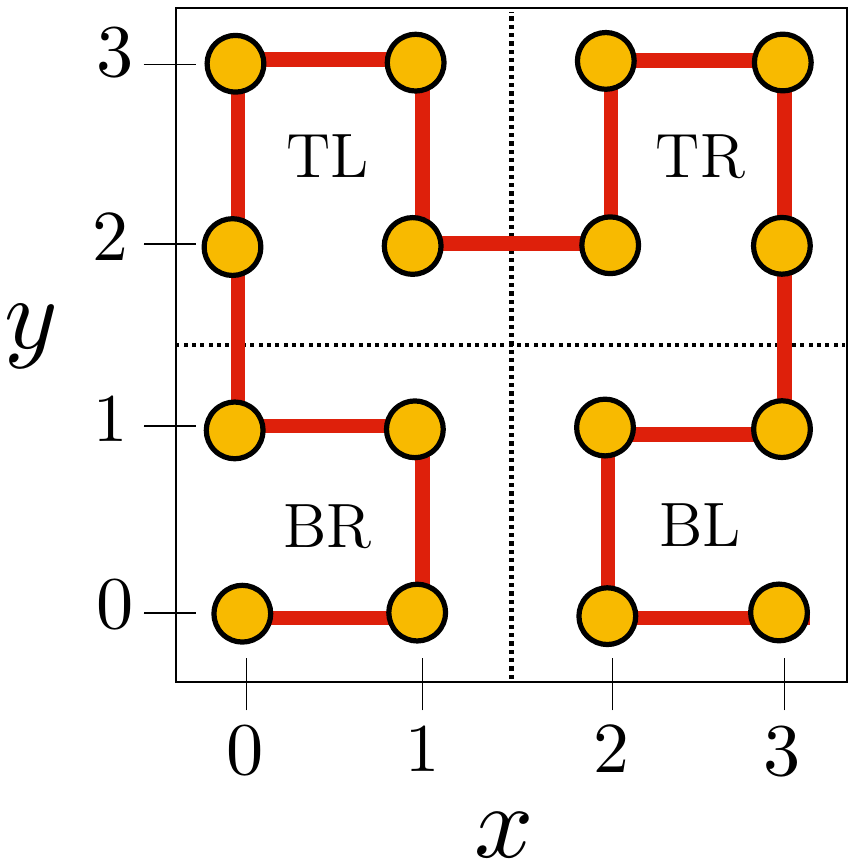}
            \label{fig:Hilbert_4}
        \end{subfigure}
        \begin{subfigure}[b]{0.48\columnwidth}   
            \centering 
            \captionsetup{position=top,singlelinecheck=off,justification=raggedright}
            \caption{}
            \includegraphics[width=\columnwidth]{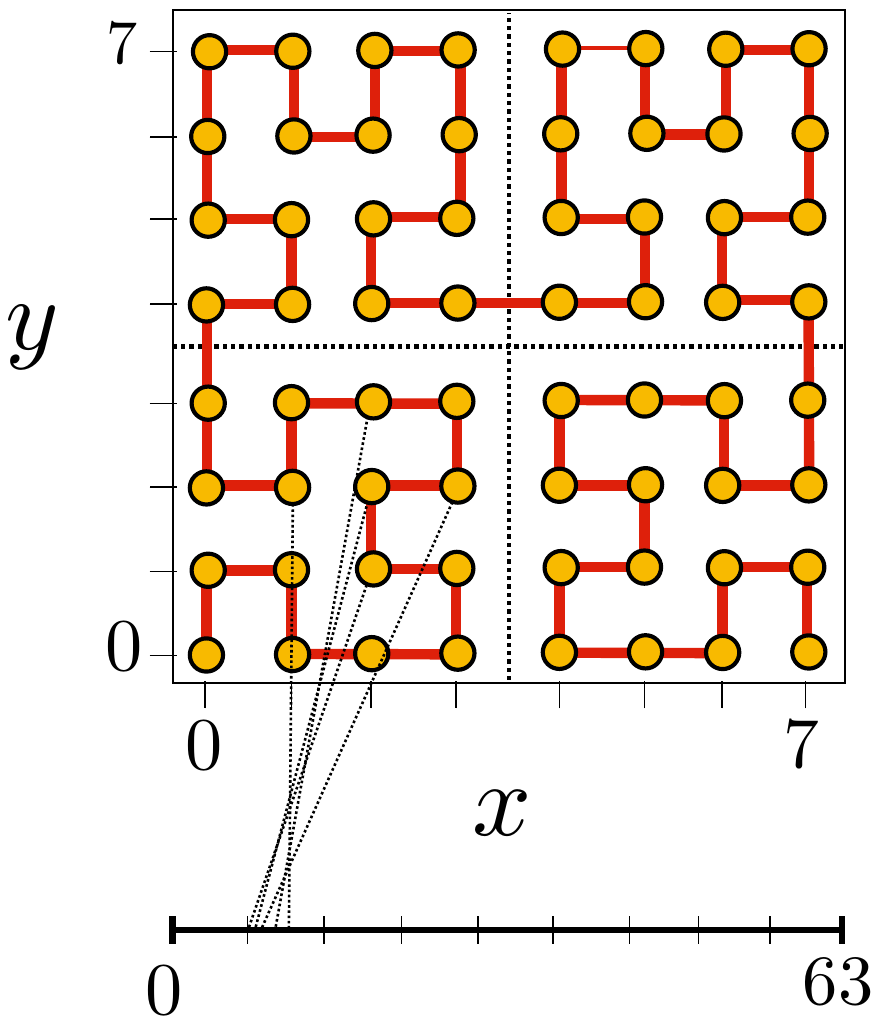}
            \label{fig:Hilbert_8}
        \end{subfigure}
        \hfill
        \begin{subfigure}[b]{0.48\columnwidth}   
            \centering
            \captionsetup{position=top,singlelinecheck=off,justification=raggedright}
             \caption{} 
            \includegraphics[width=\columnwidth]{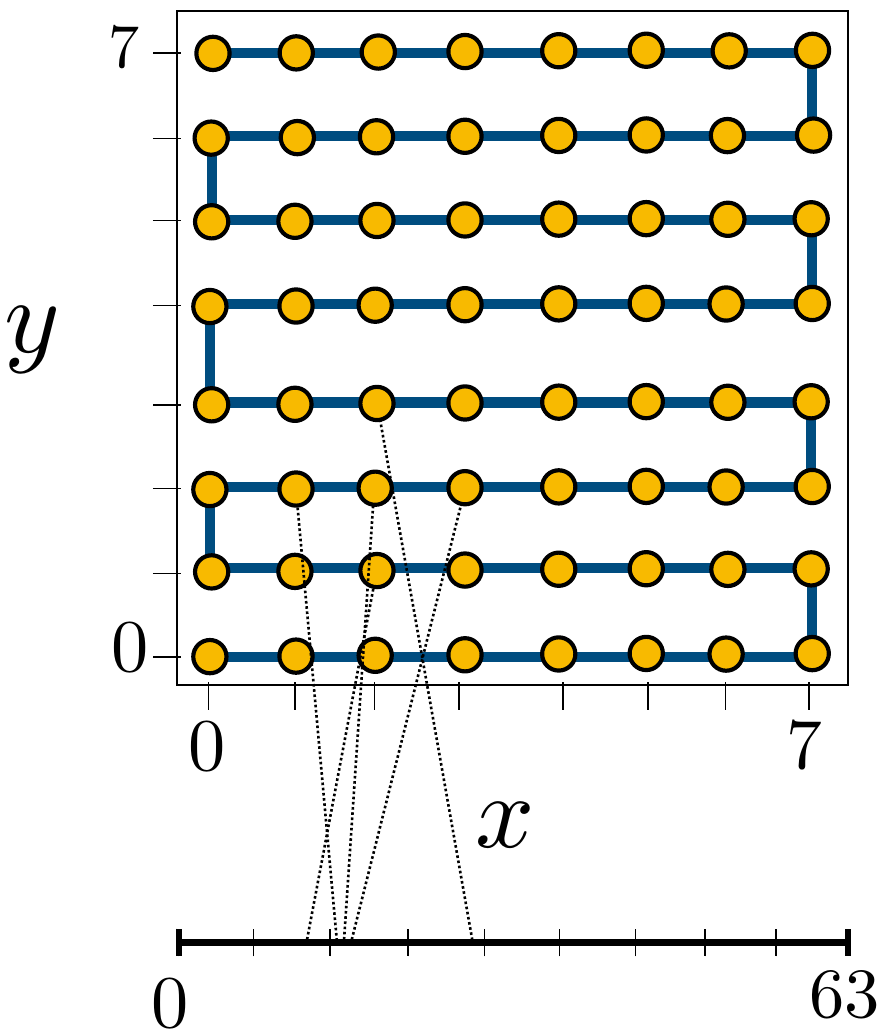}
            \label{fig:Snake_8}
        \end{subfigure}
        \captionsetup{justification=centerlast}
        \caption{(a) Hilbert curve for $n=2$, (b) for $n=4$, (c) for $n=8$. (d) Snake curve for $n=8$. In (c) and (d) it is also shown the 
projection of a generic site of the 2D lattice (within a given quadrant) and its four nearest neighbours onto the one-dimensional chain.}
        \label{fig:hilbert_and_snake_curve}
\end{figure}

The {{Hilbert curve}} is a {fractal-like self-similar} space-filling curve described for the first time in 1891 by the mathematician David Hilbert \cite{Hilbert_Curve1}. Among other interesting properties, it allows to create a mapping between a 1D and a 2D space by keeping and preserving the locality: this implies that two points that are close to each other in the 1D space are close to each other also after folding on the 2D space. The converse is not always strictly true, as unavoidable when passing 
from two-dimensions to one-dimension. However, even in this case, the curve shows a tendency to preserve the locality as much as possible {{\cite{doi:10.1080/02693799008941526, 10.1145/93605.98742, 908985}
which makes it a valuable tool for several applications in computer science and bioinformatics \cite{10.1007/978-3-540-74553-2_1, LEMIRE20112550, Hilbert_bioinformatics}.
The basic element of the Hilbert curve is 
 obtained  by connecting the elements of a $2\times2$ lattice starting  from the bottom-left (BL) to the bottom-right (BR) corner  as shown in Fig. \ref{fig:Hilbert_2}.  From this we can now construct the $n = 4$ Hilbert curve  via the following procedure: draw the $n = 2$ Hilbert curve}} into the top-left  (TL) an top-right (TR) quadrants, while rotating it clockwise and counterclockwise by 90 degrees in the BL and BR quadrants respectively. Then, join this different curve replicas, ending up with the $n = 4$ curve shown in Fig. \ref{fig:Hilbert_4}.
The generalization to the level $n$ for covering the $n \times n$ lattice 
is then straightforward: the $2n$ curve is obtained by drawing the $n$ curve in the four main quadrants of the $2n \times 2n$ lattice and by applying the just mentioned rules for the rotations in the BL and BR quadrants. In Fig. \ref{fig:Hilbert_8}, by way of example, we report the Hilbert 
curve for the $8 \times 8 $ lattice. 

In the context of numerical simulations with TN algorithms, it would be extremely useful to map a 2D lattice into a 1D chain in a way that preserves locality at most, avoiding long-range interactions of sites that are very far apart from each other. However, it is straightforward to prove that it is impossible to map a $d$-dimensional lattice into a $d'$-dimensional one, with $d' < d$, so that two neighboring sites in the $d$-lattice are always close together in the $d'$-lattice. Let us to consider, for instance, nearest neighbour sites: each point of the $d$-dimensional square 
lattice has $2d$ nearest neighbours, while only $2d'$ nearest neighbours in the $d'$-dimensional one. Thus, after the mapping, there will be at least $2(d-d')$ sites placed at a distance larger than one unit. For the specific case of a 2D square lattice and a 1D chain, this implies that the optimal solution would be a mapping such that two out of four nearest neighbours in the 2D lattice remain nearest neighbours in the 1D chain. 
If we strictly follow this argument, the best choice would be the frequently used snake curve, shown in Fig. \ref{fig:Snake_8}. However, in this way, the other two nearest neighbours of a generic site of the 2D lattice would  be rather far away from each other in the one-dimensional setup, with their distance increasing up to $2n$. If our goal is to preserve the locality also for these two other nearest neighbours, the curve that tends to globally satisfy these constraints, outperforming other types of mapping, is exactly the {{Hilbert curve. In order}} to get a general idea of this property, we report in Fig. \ref{fig:Hilbert_8} the Hilbert mapping of a generic site of the 2D lattice (within a given quadrant) 
and its four nearest neighbours: it is clearly visible that all these points that are close on the two-dimensional plane remain fairly close on the one-dimensional chain. This property, on the contrary, does not hold true for the standard snake mapping, where, as expected, two out of four nearest neighbours are almost always mapped at large distance, as shown in Fig. \ref{fig:Snake_8}. The only regions in which the Hilbert mapping does not preserve the locality are the inner borders of the main quadrants. In this 
case, nearest neighbor sites belonging to two adjacent quadrants turn out 
to be mapped quite far away from each other.

\section{2D Quantum Ising Model}\label{sec:quantum_ising_model}

In order to test the capabilities of the {{Hilbert curve mapping}} for our TN algorithms, we consider the 2D quantum Ising model in presence of a uniform transverse magnetic field along the $z$-direction \cite{2013LNP862S}. For a $n \times n$ square lattice, the Hamiltonian reads
\begin{equation}\label{eq:Ising_Hamiltonian}
H = J \sum_{\left < i,j \right >} \sigma_{i}^x \sigma_{j}^x + \lambda \sum_i \sigma_{i}^z\;, 
\end{equation}
where $\left < i, j \right >$ refers to the nearest neighbour sites and $\sigma_{i}^{\alpha}$ is the Pauli matrix in the $\alpha$-direction defined on the site~$i$. The coefficient $J$ represents the strength of the interaction between nearest neighbour spins along the $x$-axis, whereas $\lambda$ determines the strength of the external magnetic field along the $z$-axis. In the following, we consider the antiferromagnetic (AF) scenario 
for $J>0$, and, without loss of generality, we fix the energy scale by setting~$J=1$.

The two terms of the Hamiltonian in Eq. \eqref{eq:Ising_Hamiltonian} do not commute since $\left [ \sigma_i^{x}, \sigma_i^{z} \right ] \ne 0$. Thus, there exist quantum fluctuations in the system that can be tuned through the transverse field term. For $\lambda=0$, the ground state shows a 
complete antiferromagnetic order along the $x$-axis, producing a spontaneous staggered magnetization $M_s = \frac{1}{n^2} \sum_i \left < GS | \zeta_i \sigma_i^x |  GS \right > \ne 0$, in which $\zeta_i =\pm 1$ depending on the parity of the site $i$. In the limit of $| \lambda | \to \infty$, on the contrary, each spin align with the magnetic field along the $z$-axis, and, as a result, the antiferromagnetic order along the $x$-direction vanishes, i.e., $M_s = 0$. Between these two regimes, the system is expected to undergo a quantum phase transition at finite $| \lambda |$. In particular, we expect a second-order phase transition at $| \lambda_c | \approx 3.1$ \cite{PhysRevB.17.1429, Li_2018}.
\begin{figure}
   \includegraphics[width=0.49\textwidth]{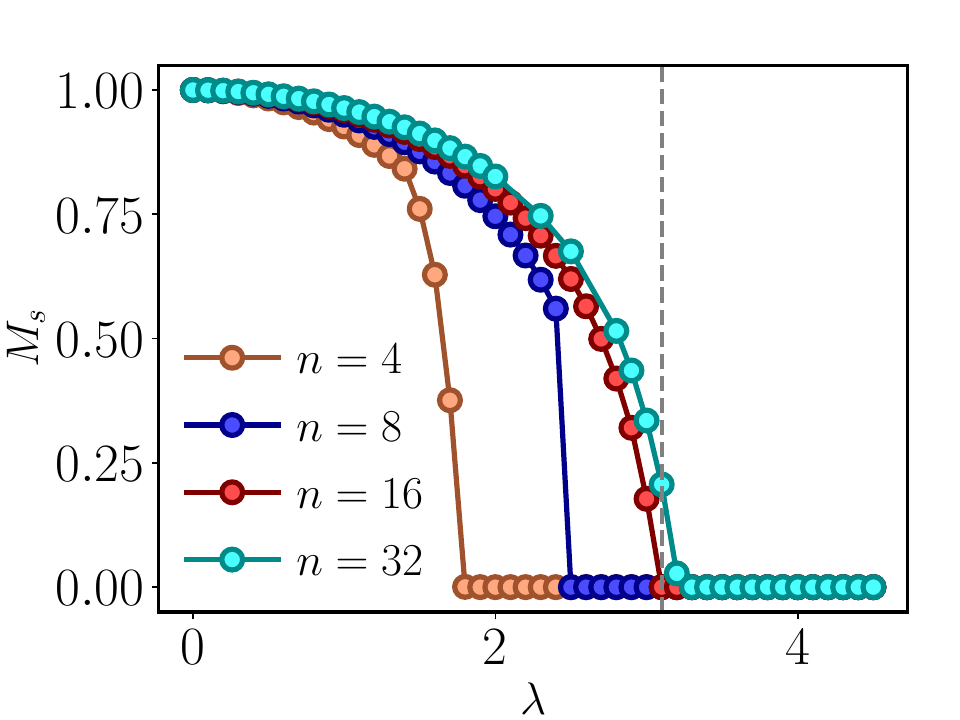}
\captionsetup{justification=centerlast}
\caption{\label{fig:der_2}
Staggered magnetization computed with TTN at OBC for different $n$ as a function of $\lambda$ and $m=25$. We observe that the magnetization drops to zero close to the expected critical point. The gray line indicates the expected position of the critical point. 
}
\end{figure} 
We apply the MPS and the TTN algorithms to numerically compute the ground 
state of the Hamiltonian in Eq. \eqref{eq:Ising_Hamiltonian} at zero temperature. We have preliminarily checked that our simulations capture the phase transition point by computing the magnetization with 
the TTN algorithm, as shown in Fig. \ref{fig:der_2} for $n=4,8,16,32$ at OBC: As expected, the magnetization drops from one to zero  close to the transition point, whose expected position is indicated by the dashed gray line \cite{PhysRevB.17.1429, Li_2018}. If not differently specified, all the results shown in the following are obtained at OBC.

\section{Numerical results}\label{sec:num_res}
Hereafter we compare  the results obtained by exploiting the Hilbert and the snake mapping with both the MPS and TTN wave function representations. {\color{black} For more details about the algorithms and their implementation see Appendix \ref{app:numerical_details}}. 
The Hilbert and the snake curves introduced above define two mappings $\mathcal{M}= \mathcal{H}, \mathcal{S}$ through which, from Eq. (\ref{eq:Ising_Hamiltonian}), we obtain two different long-range, one-dimensional Hamiltonians
\begin{equation}\label{eq:Id_ham_map}
H_{\mathcal{M}}=\sum_{<\mu,\nu>_{\mathcal{M}}}\sigma^x_\mu\sigma_\nu^x + \lambda \sum_\mu \sigma_{\mu}^z\,,
\end{equation}
where $<,>_{\mathcal{M}}$ indicates the sites mapped from nearest-neighbor coordinates in the 2D lattice according to the mapping  $\mathcal{M}$.
\begin{figure*}  
   \includegraphics[width=0.9\textwidth]{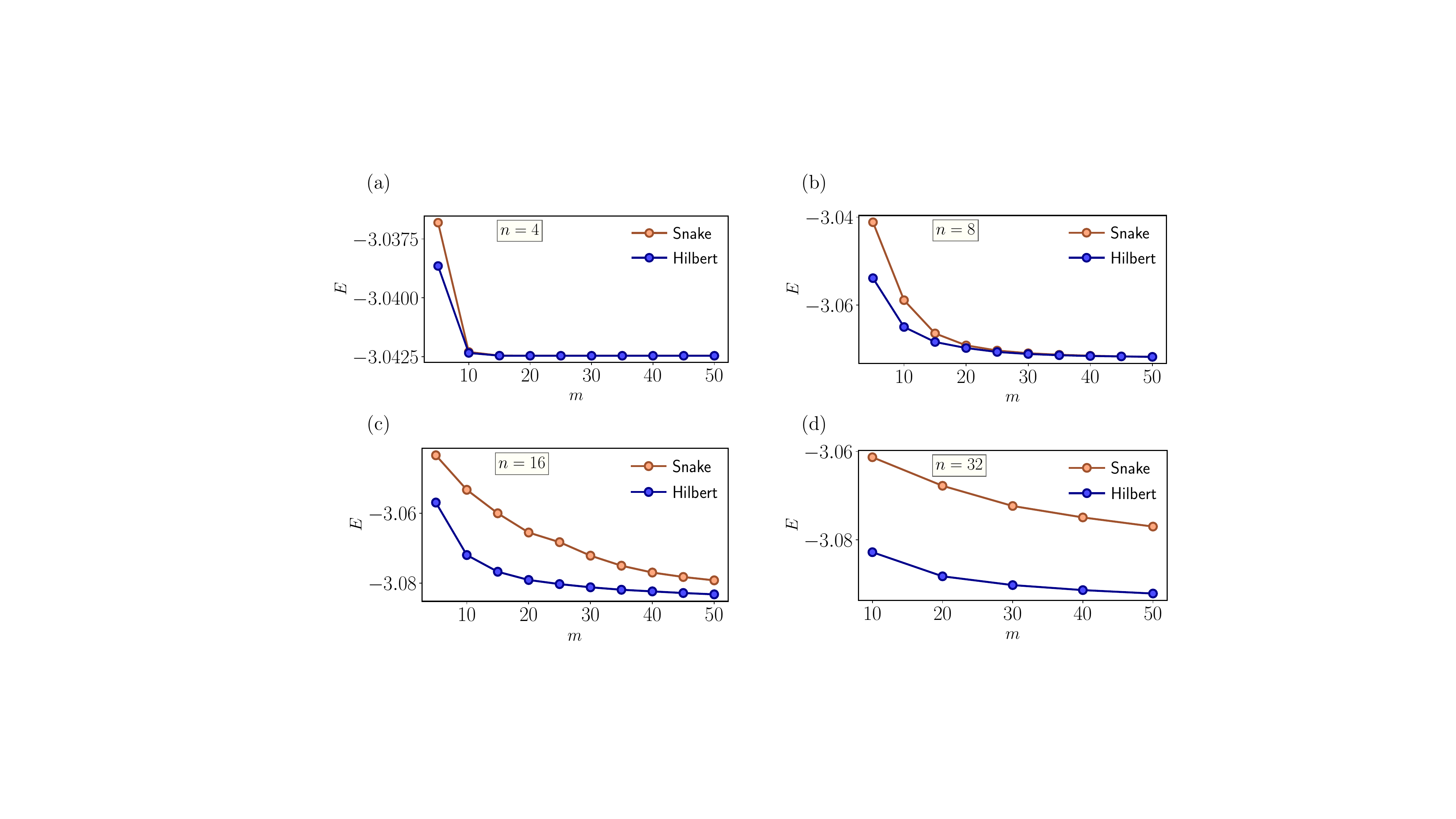}
   \captionsetup{justification=centerlast}
\caption{\label{fig:energy_per_bond_l_2.9}
{{Ground state energy density for the 2D Ising Hamiltonian~(\ref{eq:Ising_Hamiltonian})
 of a MPS on Hilbert and snake mappings, evaluated for $n=4,8,16,32$ as a function of the bond dimensions $m$ with OBC. Note how the energy difference due to the use of the Hilbert with respect to the snake one increases with the size. Here  $\lambda=2.9$ and $J=1$}}.}
\end{figure*} 
In order to estimate the performance of a simulation  with a given mapping and representation, we compute the ground state energy density $E_{\mathcal{H},\mathcal{S}}$ and the  difference $\Delta E=E_\mathcal{S} -E_\mathcal{H}$, with both the MPS and TTN algorithms.
We focus our analysis on the case $\lambda=2.9$ because, being close to 
criticality, it offers the possibility to test the two mappings in a case 
in which numerical simulations are expected to be more involving due to the presence of large correlation length. Moreover, we have numerically verified that this is the point where $\Delta E$ is larger with both MPS and TTN representation, making our analysis clearer (for details see Appendix \ref{app:app_DE}).

\begin{figure*}[t!]
   \includegraphics[width=0.9\textwidth]{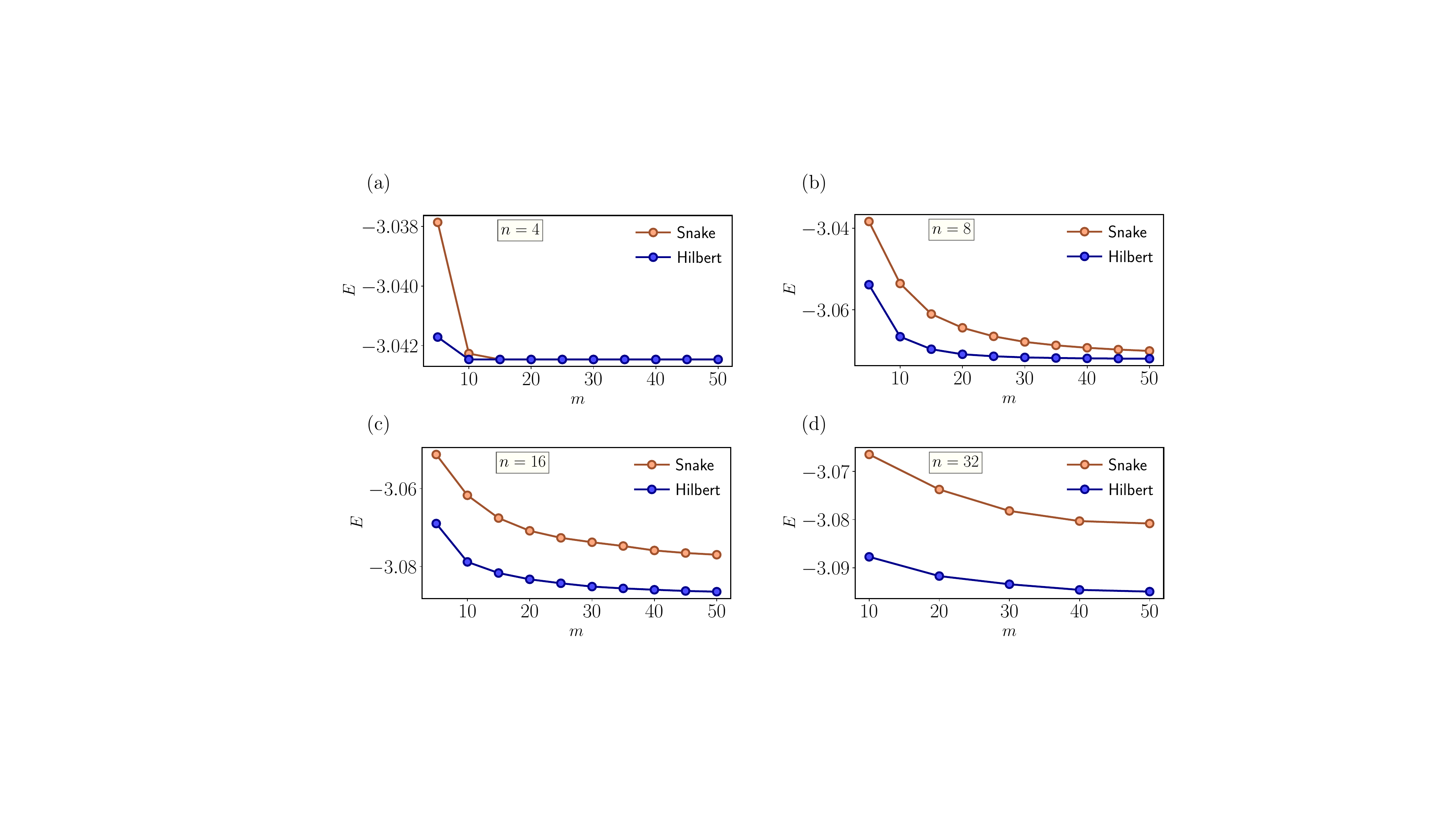}
\captionsetup{justification=centerlast}
\caption{\label{fig:gs_en_2.9}
{{Ground state energy density for the 2D Ising Hamiltonian~(\ref{eq:Ising_Hamiltonian})
computed with TTN for $n=4,8,16,32$ as a function of the bond dimensions $m$ for OBC.
Here $\lambda=2.9$ and $J=1$.  As for the MPS case the energy difference due to the use of the Hilbert with respect to the snake one increases with the size.}}
}
\end{figure*}

First, we show the results obtained by employing the MPS Ansatz for computing the Hamiltonian ground state. 
We  perform simulations for the two effective Hamiltonians $H_\mathcal{H}$ and $H_\mathcal{S}$ for $n=4,\ 8,\ 16,\ 32$, {{for $\lambda=2.9$}}, and measure the ground state energy density {{(i.e. the energy per site)}} as a function of the bond dimension $m$ (see Fig. \ref{fig:energy_per_bond_l_2.9}). In each panel, for both mappings we observe that the ground state energy {{density}}  decreases with the increase of the bond dimension, converging to an asymptotic energy value. 
We notice that a clear improvement is achieved when the Hilbert curve is adopted, with the energy difference increasing as the system size grows. 
{{Analogous results are observed in Fig.~\ref{fig:gs_en_2.9} which have been obtained for the same
parameter values using now the TTN Ansazt with OBC: also in this case the ground state energy densities  associated with the Hilbert curve mapping
are smaller than those obtained with the snake curve. This behavior emerges also for PBC, as shown in Appendix \ref{app:PBC_TTN}.
}}

\begin{figure}  
 \begin{subfigure}[b]{0.9\columnwidth}
  \captionsetup{position=top,singlelinecheck=off,justification=raggedright}
     \caption{}
         \centering
         \includegraphics[width=\columnwidth]{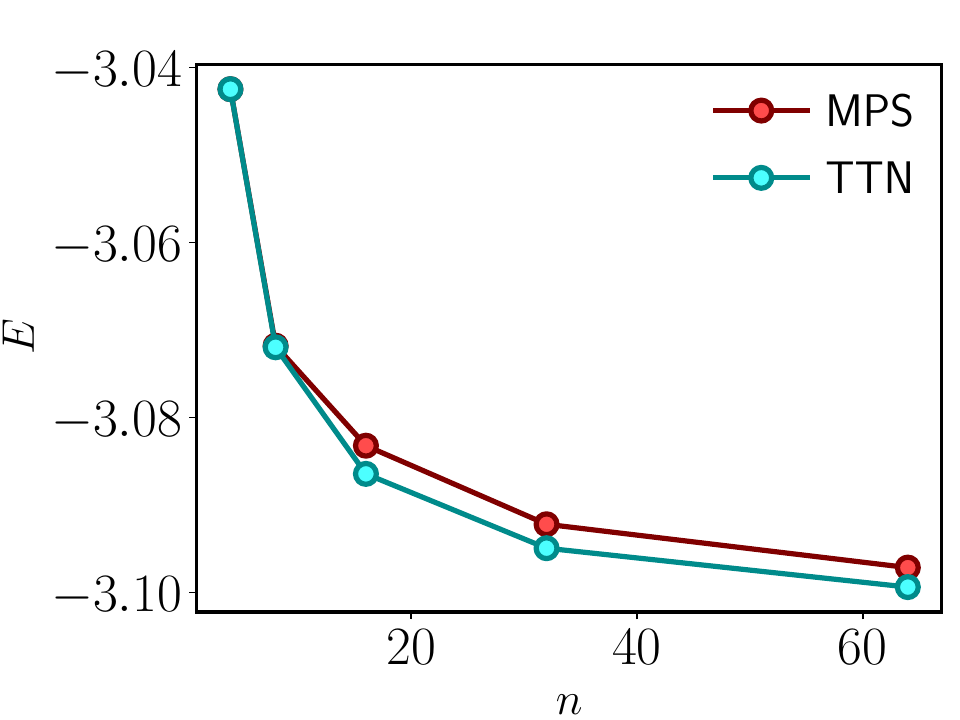}
         \label{fig:GS_MPS_TTN}
     \end{subfigure}
     
 \begin{subfigure}[b]{0.9\columnwidth}
  \captionsetup{position=top,singlelinecheck=off,justification=raggedright}
     \caption{}
         \centering
         \includegraphics[width=\columnwidth]{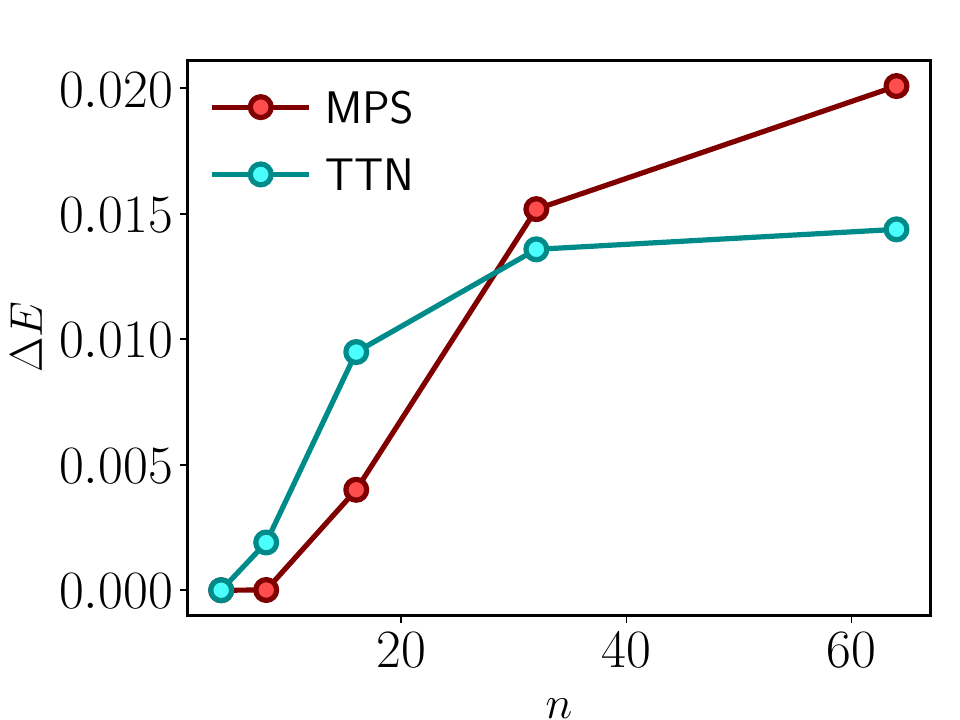}
         \label{fig:energy_gap_n_b_30__l_2.9}
     \end{subfigure}
\captionsetup{justification=centerlast}
\caption{
 (a) Ground state energy density computed with MPS and TTN at OBC by exploiting the Hilbert mapping as a function of $n$ ($\lambda=2.9$, $m=50$). (b) Energy density difference $\Delta E$ computed with the MPS and the TTN algorithms as a function of $n$. }
\end{figure}

In order to compare the results obtained with the MPS and the TTN Ansatze, we first plot the ground state energy densities for  the Hilbert curve case at $m=50$. As shown in Fig. \ref{fig:GS_MPS_TTN}, we find that the values returned by the TTN simulations are slightly lower than those found by exploiting MPS. We 
now consider the difference $\Delta E$ as a function of $n$, plotted in Fig. \ref{fig:energy_gap_n_b_30__l_2.9} up to $n=64$. We find that the Hilbert curve improves the 
energy estimation at the second decimal digit, thus outperforming the snake mapping at each $n$ both with MPS and TTN. Moreover, while at low $n$ the improvement achieved with the Hilbert mapping is larger with TTN, MPS exhibits a larger $\Delta E$ at higher values of $n$.

\begin{figure}
 \begin{tabular}{c}
                \begin{subfigure}[b]{0.97\columnwidth}
      \captionsetup{position=top,singlelinecheck=off,justification=raggedright}
      \caption{}
         \centering
          \includegraphics[width=\columnwidth]{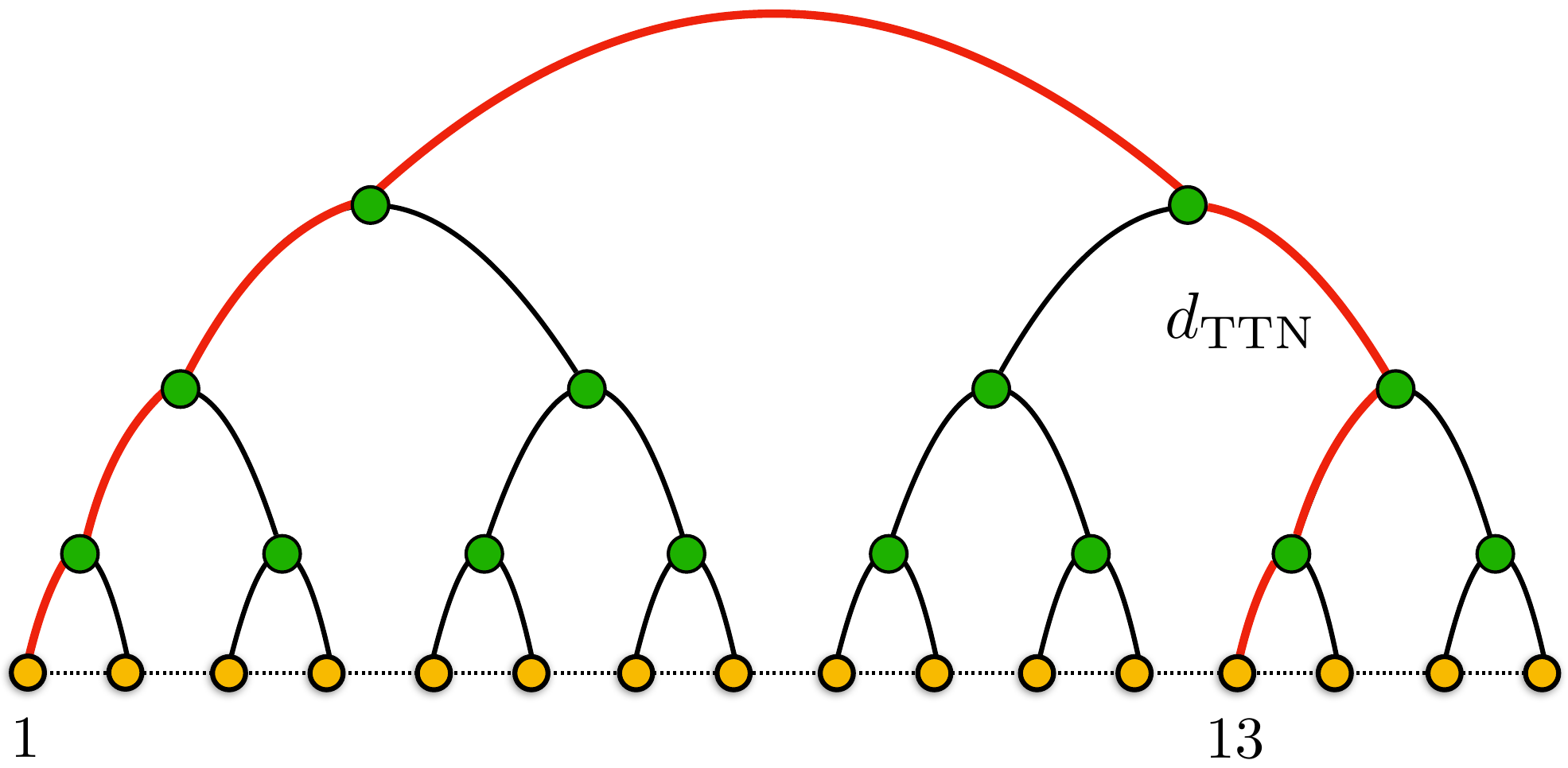}
         \label{fig:distances_TTN}
     \end{subfigure} 
     \\
                    \begin{subfigure}[b]{0.97\columnwidth}
      \captionsetup{position=top,singlelinecheck=off,justification=raggedright}
      \caption{}
         \centering
          \includegraphics[width=\columnwidth]{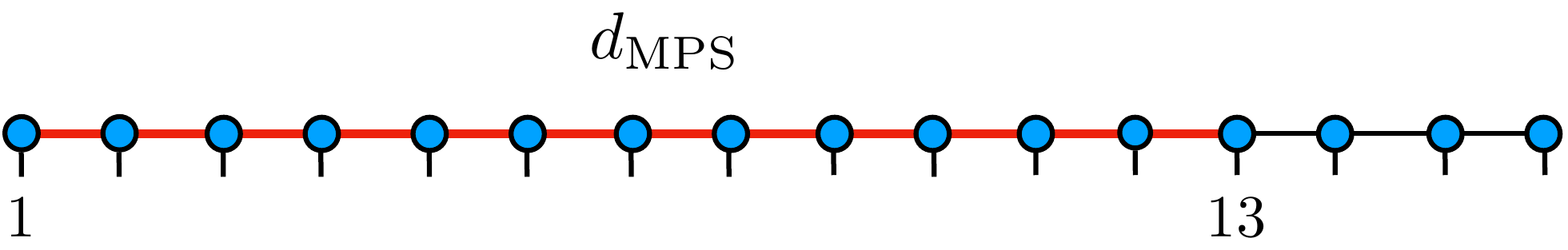}
         \label{fig:distances_MPS}
     \end{subfigure}
   \end{tabular}
\captionsetup{justification=centerlast}
\caption{\label{fig:distances_both} Distances between two lattices sites computed within (a) TTN and (b) MPS geometries. In this case, for the sites 1 and 13, $d_{\mathrm{TTN}}=7$ and $d_{\mathrm{MPS}}=12$}.
\end{figure}

{\color{black}In order to provide a qualitative explanation to our findings, let us introduce two distances $d_\mathrm{MPS}$ and $d_\mathrm{TTN}$, defined as the number of links connecting two different lattice sites within the MPS and TTN network geometries respectively, as shown in Fig. \ref{fig:distances_both}. Thus, we have considered the interaction terms of the 1D Hamiltonians $H_\mathcal{S}$ and $H_\mathcal{H}$ and we have computed the distances $d_\mathrm{MPS}$ and $d_\mathrm{TTN}$ between all the relative pairs of sites, producing four sets of distances, each of them labeled by the relative mapping and the tensor network geometry.}

\begin{figure}
 \begin{tabular}{cc}
           \begin{subfigure}[b]{0.48\columnwidth}
      \captionsetup{position=top,singlelinecheck=off,justification=raggedright}
      \caption{}
         \centering
          \includegraphics[width=\columnwidth]{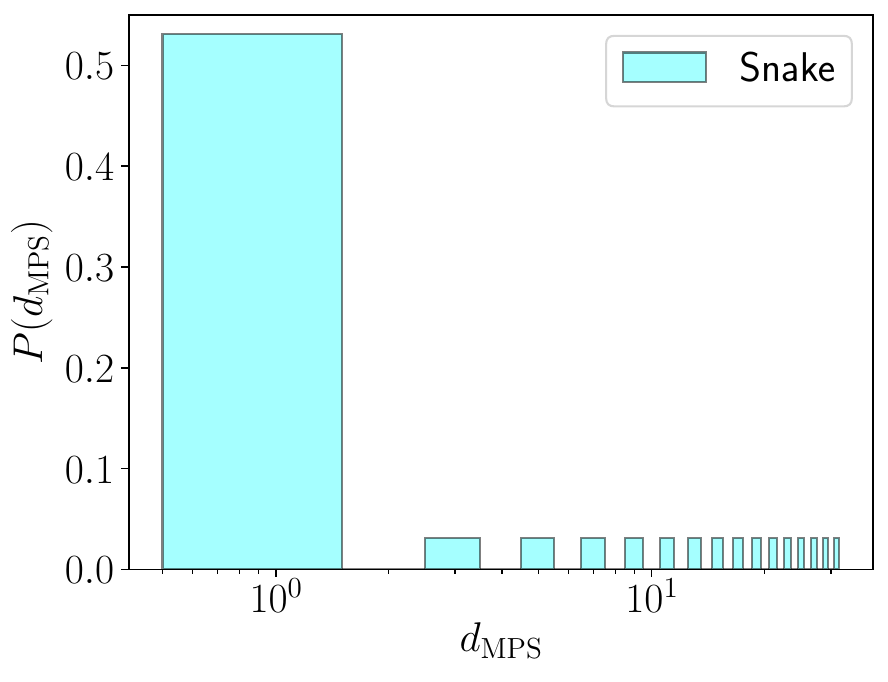}
         \label{fig:comp_snake_hilb_loga}
     \end{subfigure}&
               \begin{subfigure}[b]{0.48\columnwidth}
      \captionsetup{position=top,singlelinecheck=off,justification=raggedright}
      \caption{}
         \centering
          \includegraphics[width=\columnwidth]{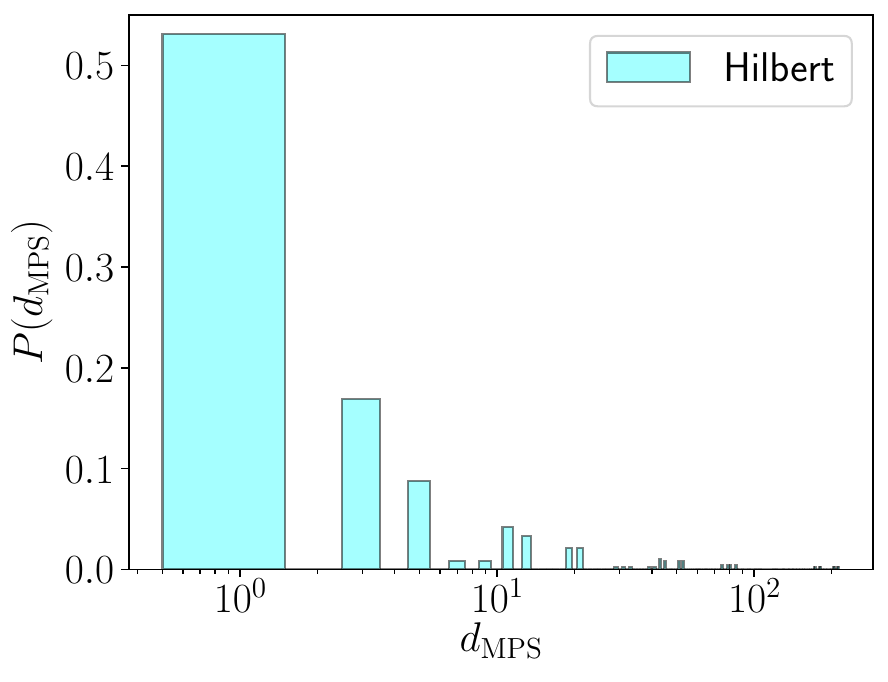}
         \label{fig:comp_snake_hilb_logb}
     \end{subfigure}\\
               \begin{subfigure}[b]{0.48\columnwidth}
      \captionsetup{position=top,singlelinecheck=off,justification=raggedright}
      \caption{}
         \centering
          \includegraphics[width=\columnwidth]{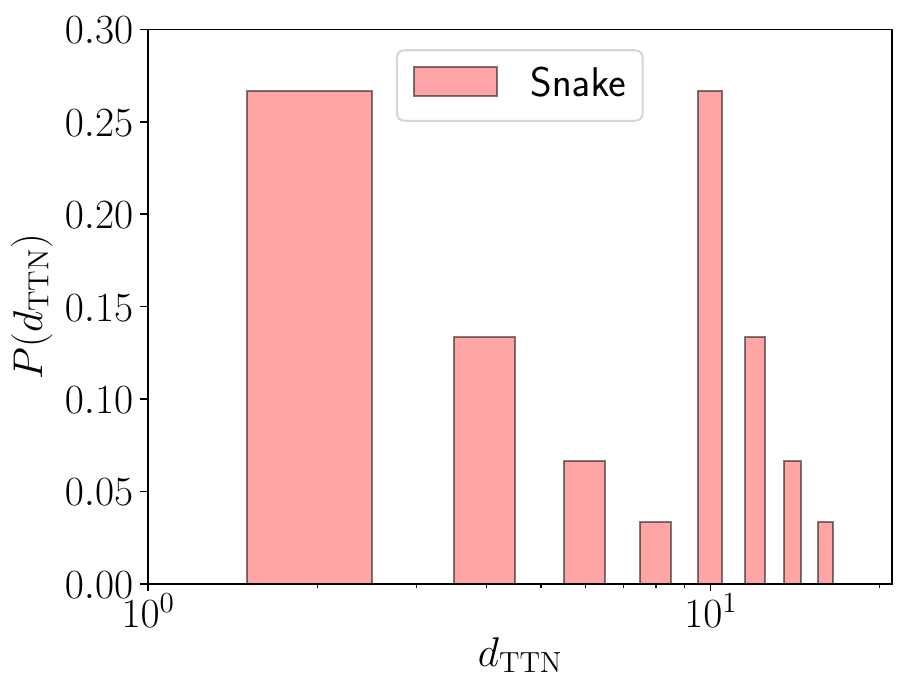}
         \label{fig:comp_snake_hilb_logc}
     \end{subfigure}&
               \begin{subfigure}[b]{0.48\columnwidth}
      \captionsetup{position=top,singlelinecheck=off,justification=raggedright}
      \caption{}
         \centering
          \includegraphics[width=\columnwidth]{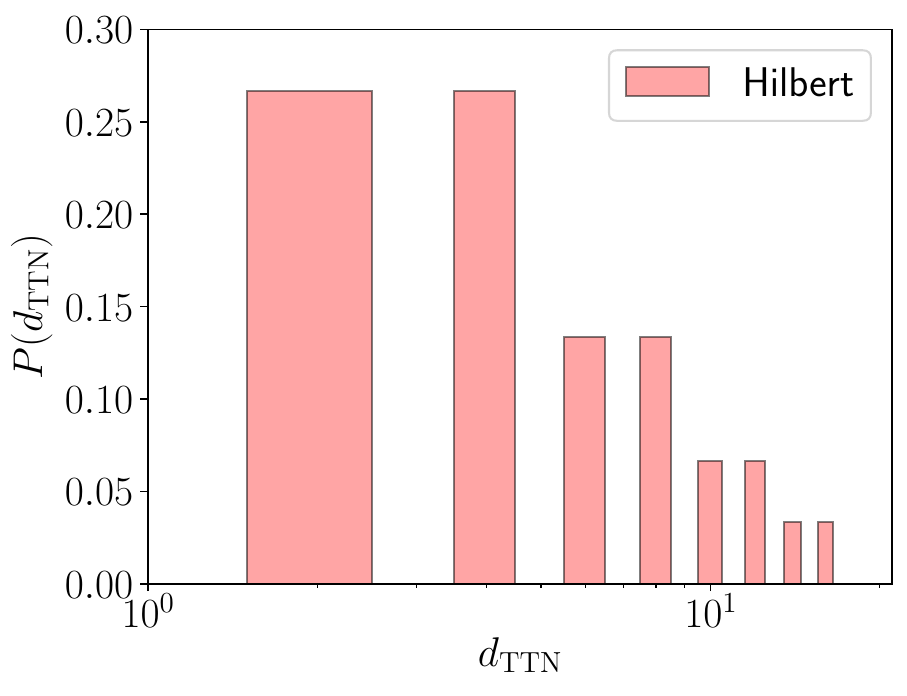}
         \label{fig:comp_snake_hilb_logd}
     \end{subfigure}
   \end{tabular}
\captionsetup{justification=centerlast}
\caption{\label{fig:comp_snake_hilb_log}
{\color{black} Distributions of the number of tensor-network links separating physically adjacent lattice sites in the 2D lattice $d_\mathrm{MPS}$ (first row) and $d_\mathrm{TTN}$ (second row) relative to the pairs of sites connected by an interaction term in the Hamiltonians $H_\mathcal{S}$ (fist column) and $H_\mathcal{H}$ (second column), computed for $n=16$.}
}
\end{figure} 

\begin{figure}
 \begin{tabular}{cc}
                \begin{subfigure}[b]{0.48\columnwidth}
      \captionsetup{position=top,singlelinecheck=off,justification=raggedright}
      \caption{}
         \centering
          \includegraphics[width=\columnwidth]{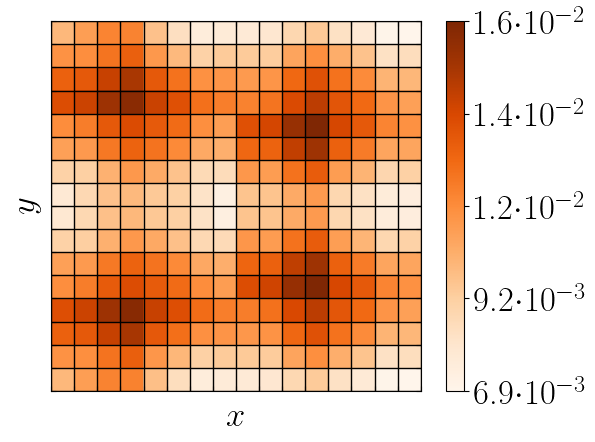}
         \label{fig:delta_magna}
     \end{subfigure}&
                    \begin{subfigure}[b]{0.48\columnwidth}
      \captionsetup{position=top,singlelinecheck=off,justification=raggedright}
      \caption{}
         \centering
          \includegraphics[width=\columnwidth]{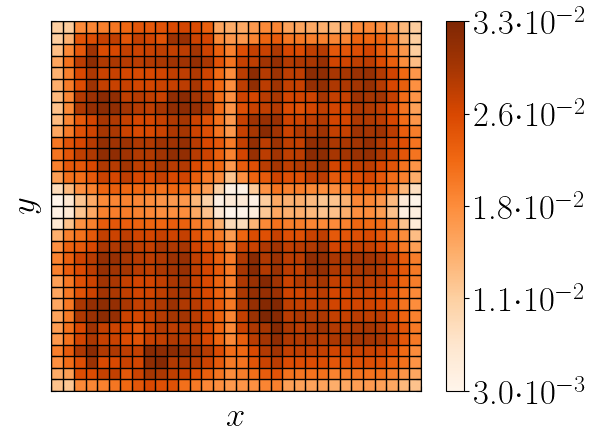}
         \label{fig:delta_magnb}
     \end{subfigure} \\
                     \begin{subfigure}[b]{0.48\columnwidth}
      \captionsetup{position=top,singlelinecheck=off,justification=raggedright}
      \caption{}
         \centering
          \includegraphics[width=.92\columnwidth]{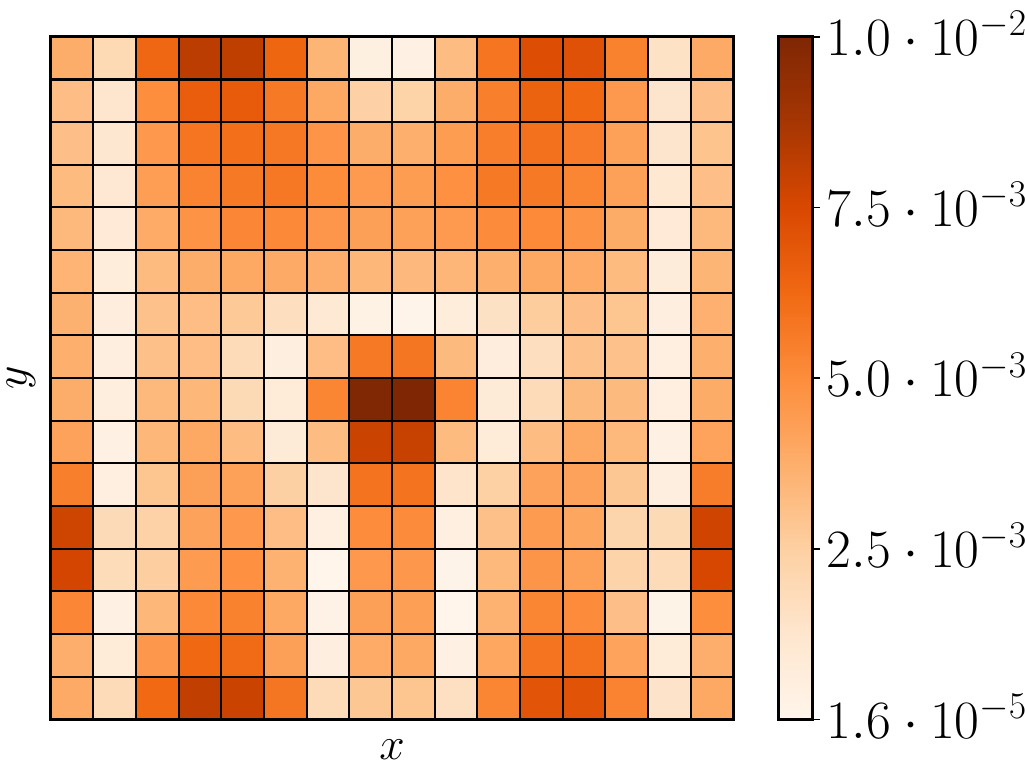}
         \label{fig:delta_magnc}
     \end{subfigure}&
                    \begin{subfigure}[b]{0.48\columnwidth}
      \captionsetup{position=top,singlelinecheck=off,justification=raggedright}
      \caption{}
         \centering
          \includegraphics[width=.92\columnwidth]{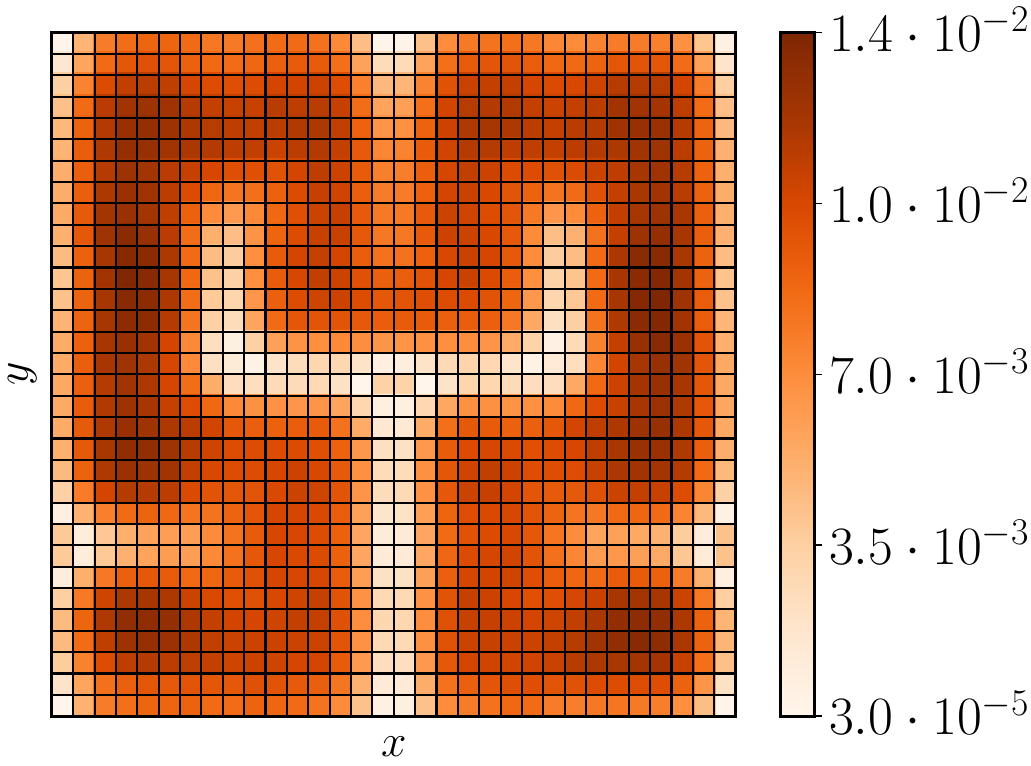}
         \label{fig:delta_magnd}
     \end{subfigure}
   \end{tabular}
\captionsetup{justification=centerlast}
\caption{\label{fig:delta_magn}
\textcolor{black}{Local magnetization difference $\Delta \sigma^z_\mu$ between the Hilbert and the snake ground states for $n=16$ (left panel) and $n=32$ (right panel) computed by using TTNs (upper row) and MPS (lower row) for $\lambda=2.9,\, m=50$ with PBC. The regions in which the expectation values are similar are those in which neither the Hilbert nor the snake curve are able to properly preserve the locality of the interactions.}
}
\end{figure}

In Fig. \ref{fig:comp_snake_hilb_log}, {\color{black} we fix $n=16$} and we plot the distributions of {\color{black} these} distances $d_\mathrm{MPS}$ and $d_\mathrm{TTN}$ for the snake (left column) and the Hilbert (right column) mappings: one can qualitatively see that the probabilities at  large distances are larger for the snake mapping than for the Hilbert one. 
 Furthermore, this feature is particularly evident in the TTN case, due to the logarithmic scaling of distances within the network. 
 It is possible to understand which regions of the lattice contribute 
to the energy improvement by studying the expectation values of the local magnetization. \textcolor{black}{We measure the local expectation values of the magnetization $\{\langle\sigma^z_\mu\rangle\}_{1\leq \mu \leq n^2}$ on both the ground states of the two Hamiltonians $H_\mathcal{H}$ and $H_\mathcal{S}$ and computing the difference $\Delta \sigma^z_\mu=|\langle\sigma^z_\mu\rangle_\mathcal{H}-\langle\sigma^z_\mu\rangle_\mathcal{S}|$, as shown in Fig. \ref{fig:delta_magn}. }
 There are regions in which the difference is larger while in others the magnetization values are similar. 
The former are  located in the bulk of the lattice main quadrants (regions delimited by black lines in Fig. \ref{fig:Hilbert_8}) where the 
Hilbert curve preserves the locality of interactions better than the snake one.
Instead, the latter are the regions close to the quadrant boundaries, where the Hilbert curve fails to preserve locality and the magnetization results are not more precise than those obtained with the snake one. 

\begin{figure}
 \begin{tabular}{cc}
               \begin{subfigure}[b]{0.48\columnwidth}
      \captionsetup{position=top,singlelinecheck=off,justification=raggedright}
      \caption{}
         \centering
          \includegraphics[width=\columnwidth]{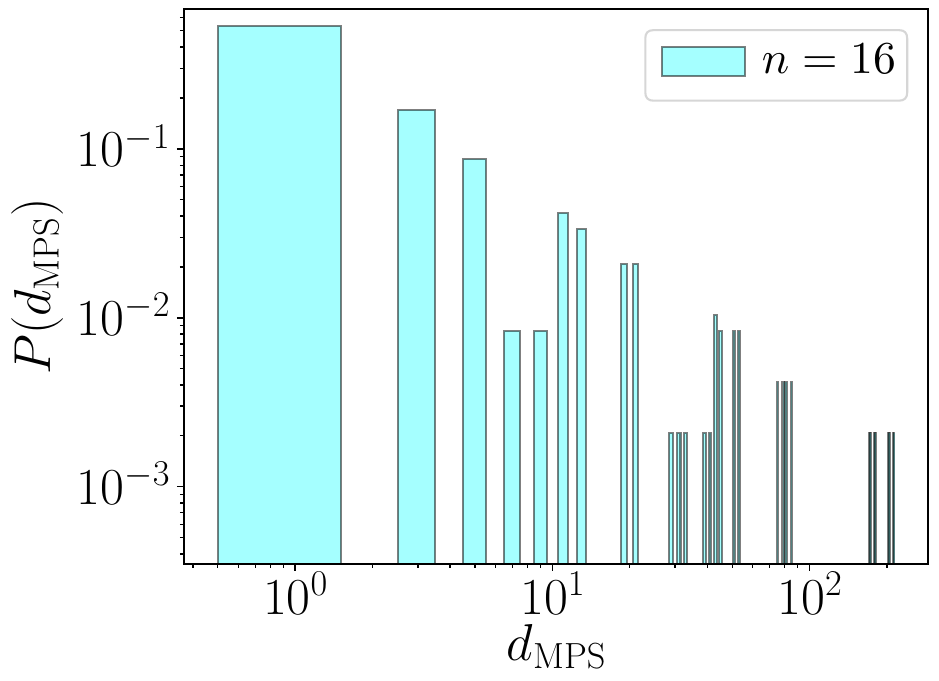}
         \label{fig:comp_MPS_TTN_loga}
     \end{subfigure}&
                   \begin{subfigure}[b]{0.48\columnwidth}
      \captionsetup{position=top,singlelinecheck=off,justification=raggedright}
      \caption{}
         \centering
          \includegraphics[width=\columnwidth]{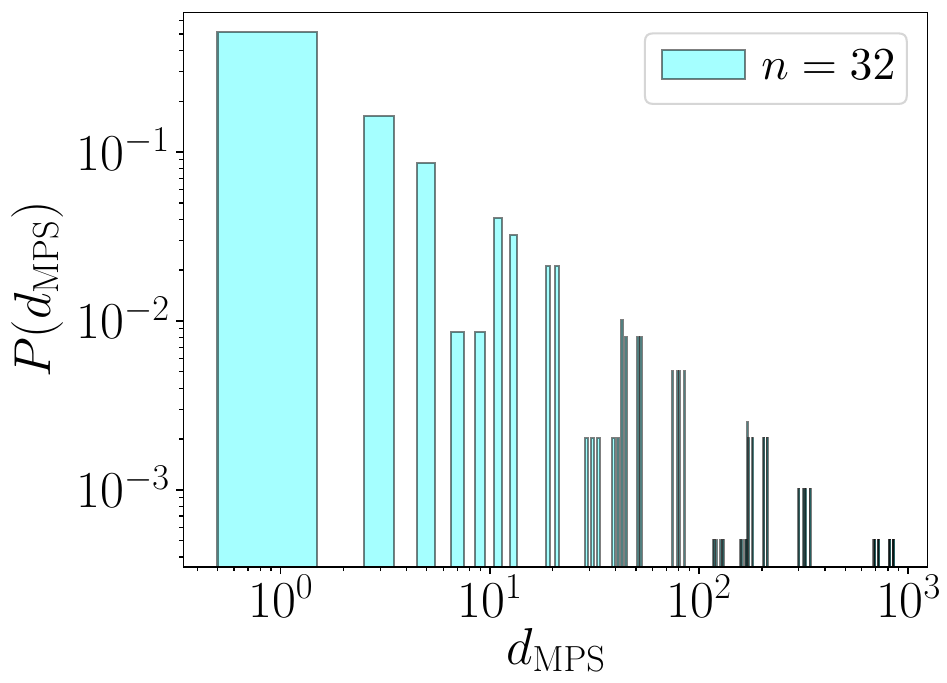}
         \label{fig:comp_MPS_TTN_logb}
     \end{subfigure}\\
                   \begin{subfigure}[b]{0.48\columnwidth}
      \captionsetup{position=top,singlelinecheck=off,justification=raggedright}
      \caption{}
         \centering
          \includegraphics[width=\columnwidth]{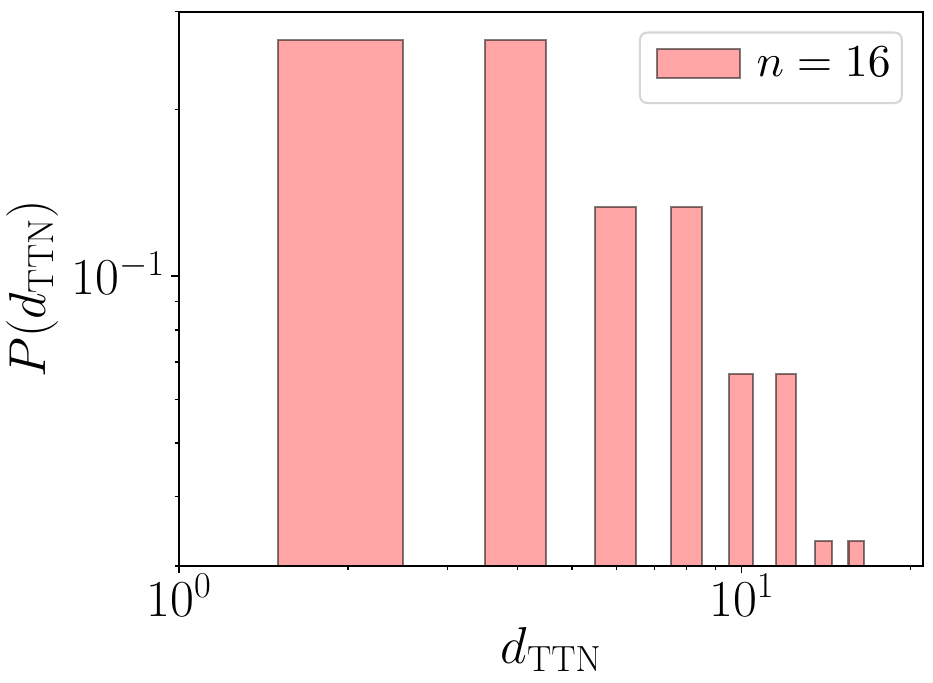}
         \label{fig:comp_MPS_TTN_logc}
     \end{subfigure}&
                   \begin{subfigure}[b]{0.48\columnwidth}
      \captionsetup{position=top,singlelinecheck=off,justification=raggedright}
      \caption{}
         \centering
          \includegraphics[width=\columnwidth]{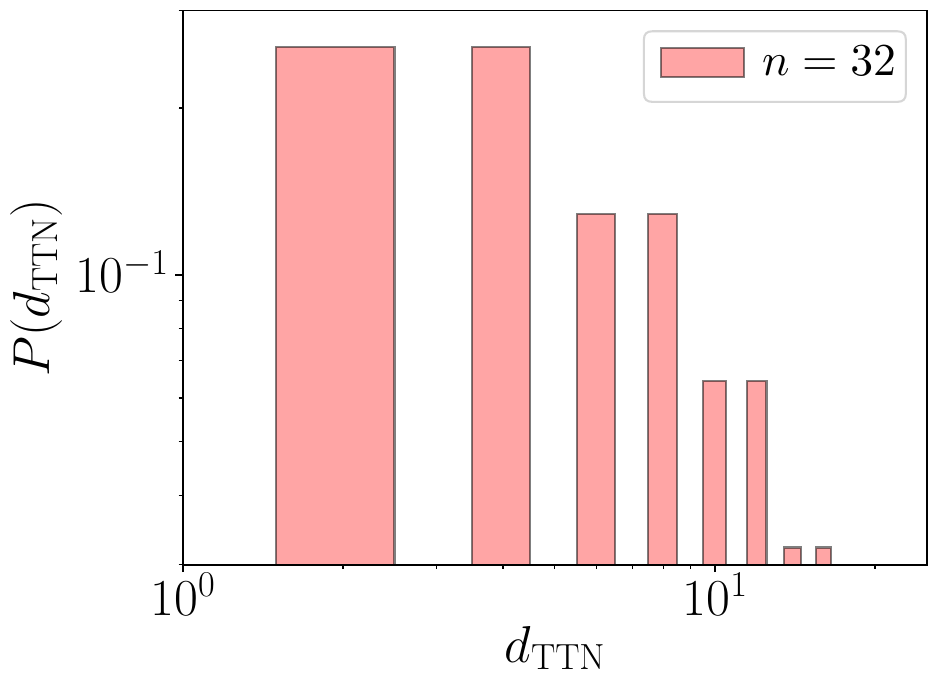}
         \label{fig:comp_MPS_TTN_logd}
     \end{subfigure}
   \end{tabular}
\captionsetup{justification=centerlast}
\caption{\label{fig:comp_MPS_TTN_log}
{\color{black} Distributions of the distances $d_\mathrm{MPS}$ (first row) and $d_\mathrm{TTN}$ (second row) relative to the pairs of sites connected by an interaction term in the Hamiltonian $H_{\mathcal{H}}$ computed for $n=16$ (first column) and $n=32$ (second column).}
}
\end{figure}

To explain, instead, why the improvement observed at large $n$ is larger for MPS with respect to the TTNs, we have checked how the fraction of interaction terms with largest distances changes when we move from $n=16$ to $n=32$. We have focused on the case of the Hilbert curve, finding that this fraction decreases faster in the MPS case than in the TTN one, as shown in Fig. \ref{fig:comp_MPS_TTN_log}. For this reason, 
we expect the improvement achievable with the MPS and the Hilbert curve is larger than the one achievable with TTNs, even though the TTN simulations are overall more accurate than the MPS ones.

\begin{figure}  
\begin{tabular}{cc}
 \begin{subfigure}[b]{0.48\columnwidth}   
            \centering
            \captionsetup{position=top,singlelinecheck=off,justification=raggedright}
             \caption{} 
            \includegraphics[width=\columnwidth]{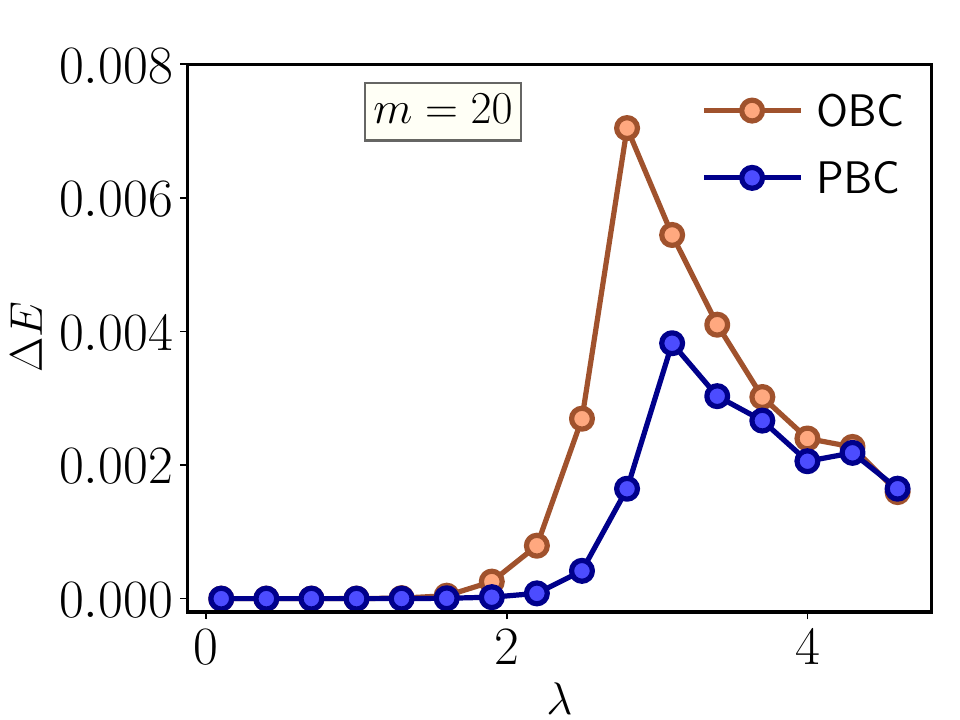}
            \label{fig:8_29}
        \end{subfigure}&
         \begin{subfigure}[b]{0.48\columnwidth}   
            \centering
            \captionsetup{position=top,singlelinecheck=off,justification=raggedright}
             \caption{} 
            \includegraphics[width=\columnwidth]{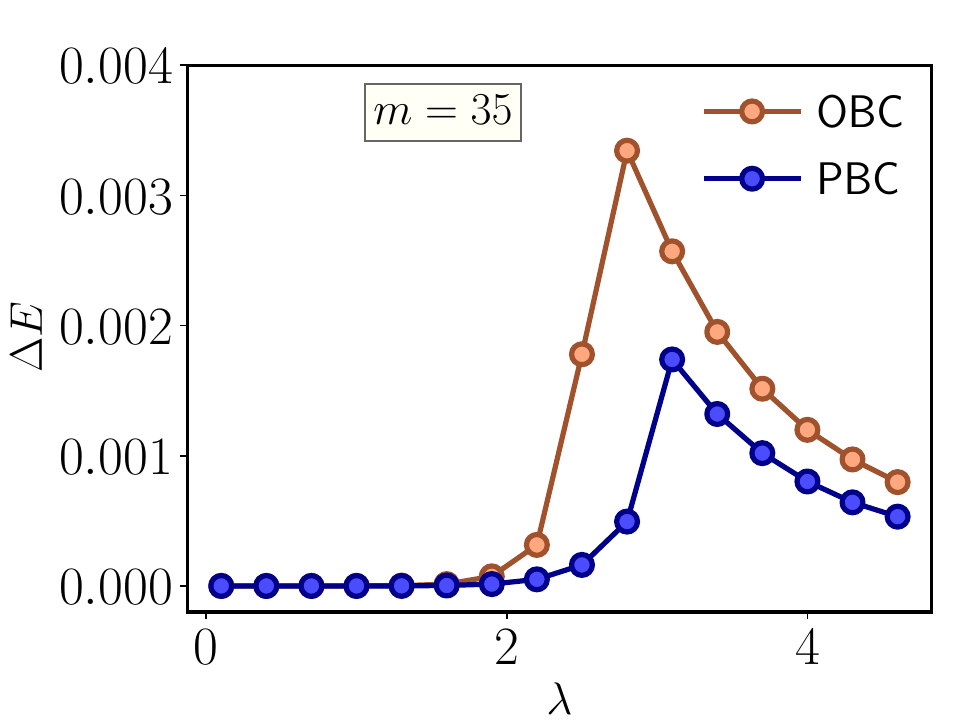}
            \label{fig:8_35}
        \end{subfigure}\\
         \begin{subfigure}[b]{0.48\columnwidth}   
            \centering
            \captionsetup{position=top,singlelinecheck=off,justification=raggedright}
             \caption{} 
            \includegraphics[width=\columnwidth]{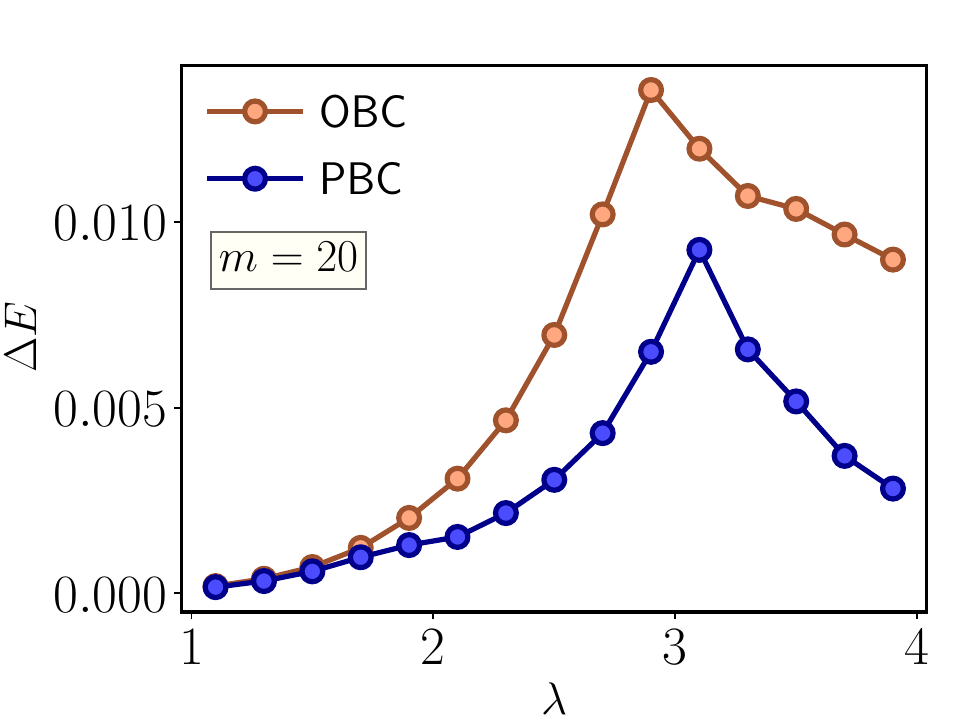}
            \label{fig:16_20}
        \end{subfigure} &
         \begin{subfigure}[b]{0.48\columnwidth}   
            \centering
            \captionsetup{position=top,singlelinecheck=off,justification=raggedright}
             \caption{} 
            \includegraphics[width=\columnwidth]{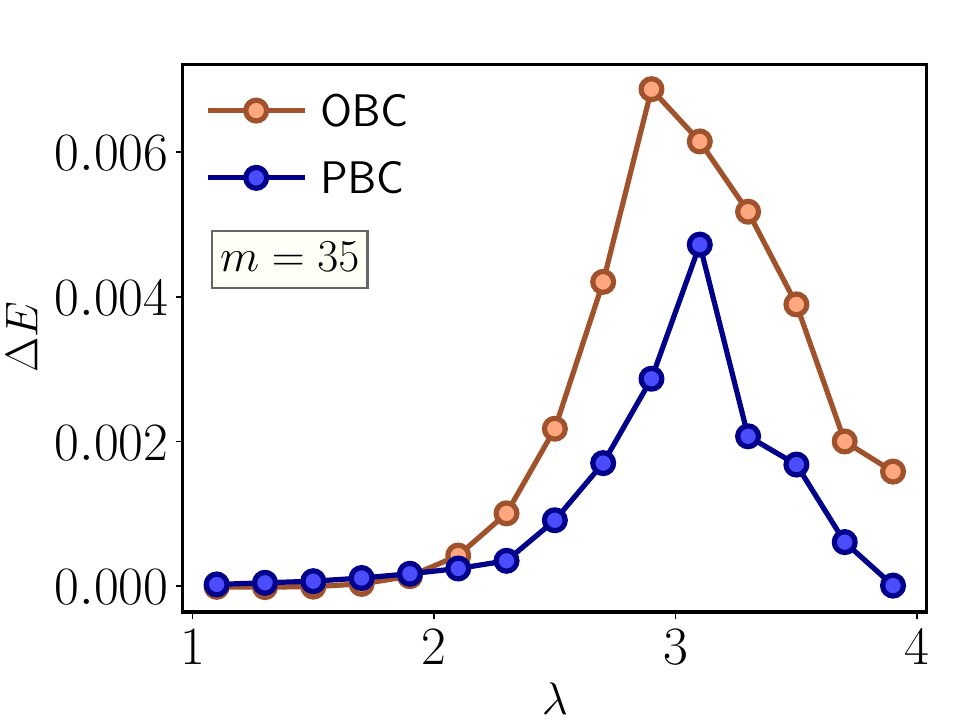}
            \label{fig:16_35}
        \end{subfigure}
   \end{tabular}
\captionsetup{justification=centerlast}
\caption{\label{fig:app_energy_gap} 
{{ Difference $\Delta E$ of the ground state energy densities}} computed with bond dimension $m=20$ (left column) and $m=35$ (right column). In the first (second) row TTN (MPS) results are shown for $n=8$ ($n=16$) with OBC and PBC.}
\end{figure} 

\begin{figure*}[t!]
   \centering
   \includegraphics[width=0.85\textwidth]{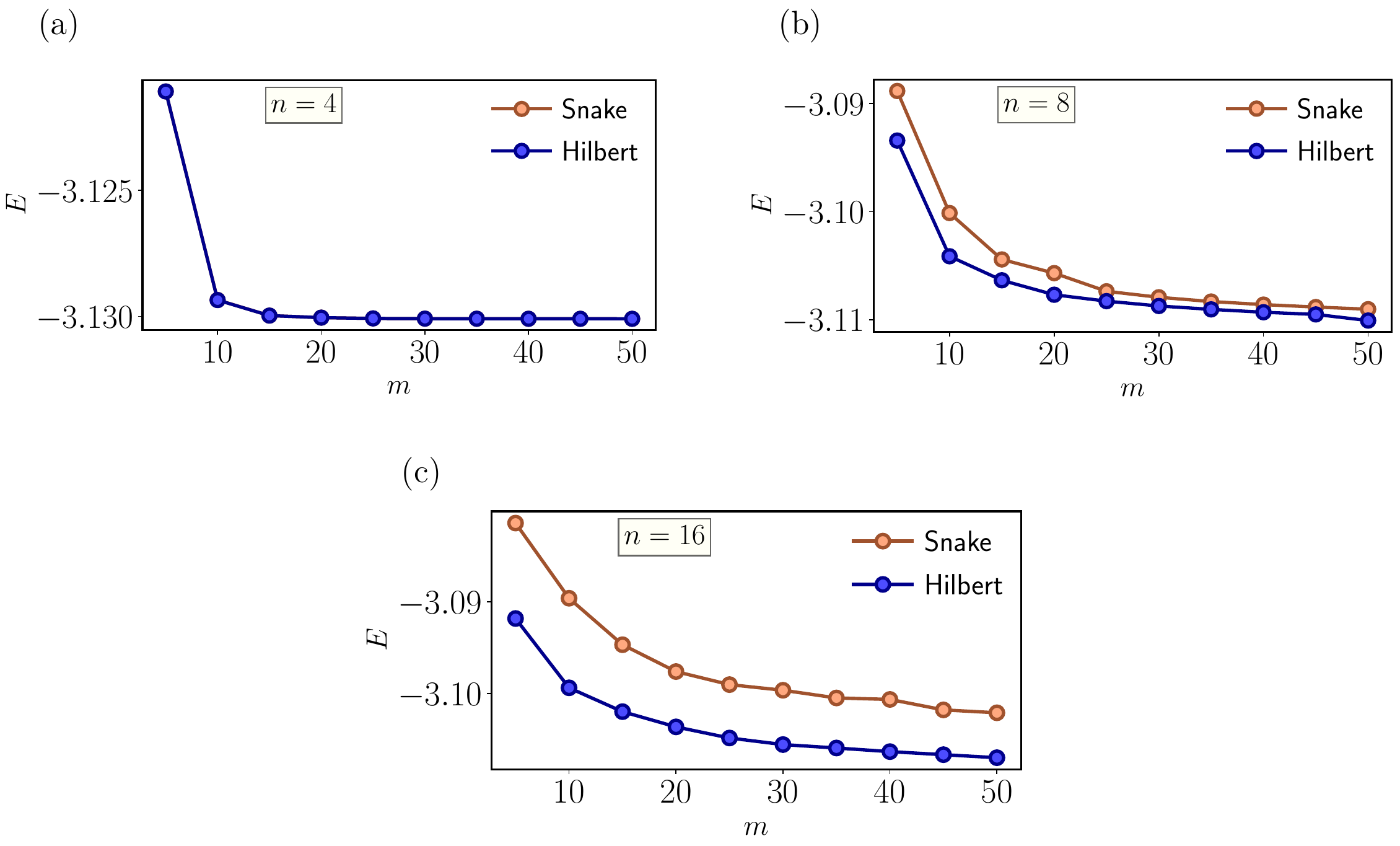}\\
\captionsetup{justification=centerlast}
\caption{\label{fig:app_PBC_E_gs_ttn}
Ground state energy {{density}} computed with TTN for $n=4,8,16$ as a function of the bond dimensions $m$ for PBC. Note how the energy difference due to the 
use of the Hilbert with respect to the snake one increases with the size, 
as found in the OBC case.
}
\end{figure*} 

\begin{figure*}[t!]
\centering
   \includegraphics[width=0.85\textwidth]{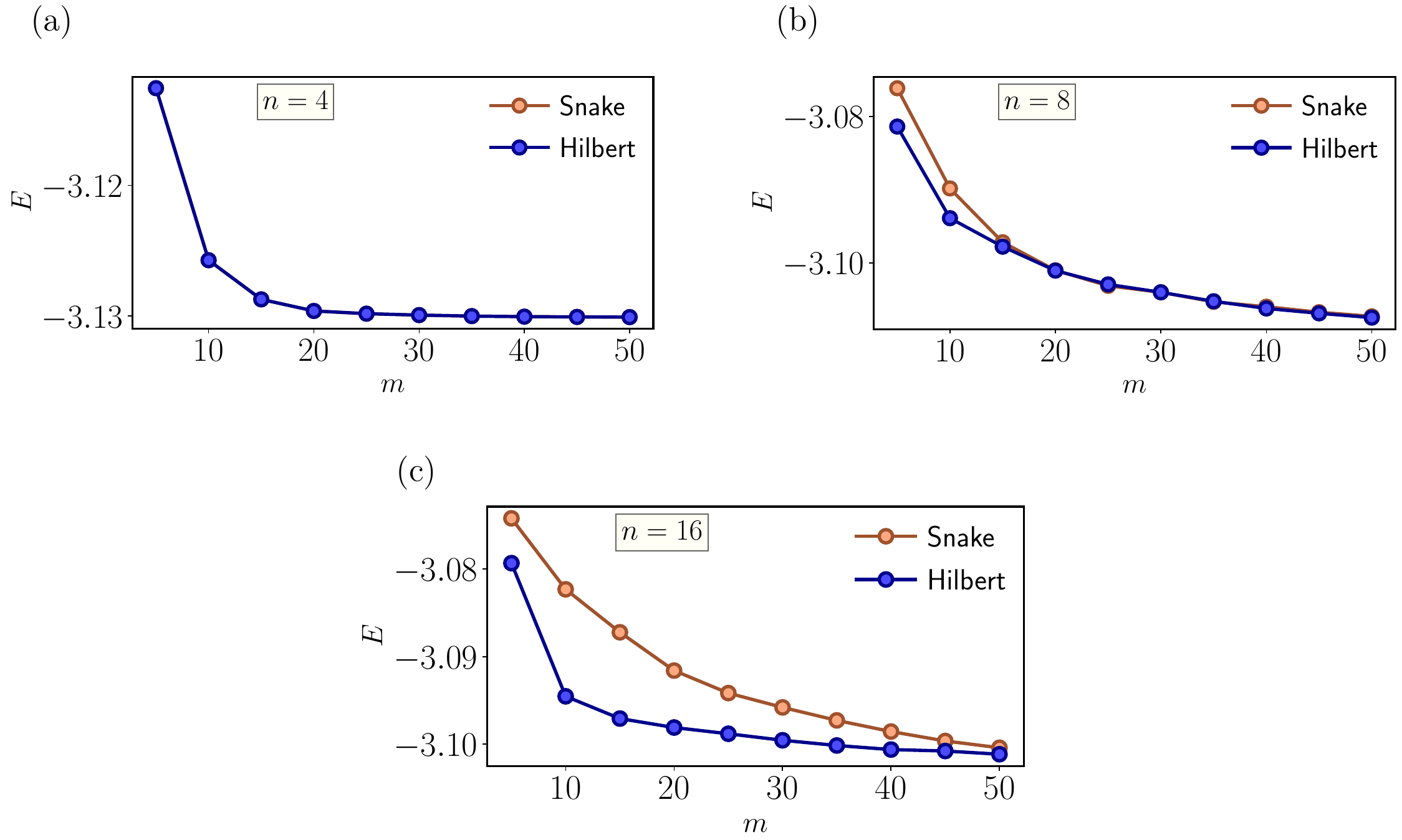}\\
\captionsetup{justification=centerlast}
\caption{\label{fig:app_PBC_E_gs_mps}
Ground state energy {{density}} computed with MPS for $n=4,8,16$ as a function of the bond dimensions $m$ for PBC. Note how the energy difference due to the 
use of the Hilbert with respect to the snake one increases with the size, 
as found in the OBC case.
}
\end{figure*} 

\section{Conclusions}\label{sec:conclusions}

In this work,  by using the MPS and the TTN algorithms, we have examined two different strategies {to study two-dimensional quantum many-body systems}. Both of them rely on the fact that it is possible to map a $d-$dimensional lattice onto a one-dimensional one, by exploiting a suitable space-filling curve. 
 In particular, we have carried out a comparison between two possible mappings, one based on the Hilbert curve and one on the snake ordering, finding a clear advantage in the use of the former, both in the MPS and in TTN approach. 
While the locality properties of the Hilbert ordering are directly mapped 
in the MPS chain, in the case of TTNs they are enhanced by the logarithmic scaling of distances within the TTN structure.
By recalling that the standard way to use TTNs for two-dimensional systems relies directly on the construction of a binary 2D TTN \cite{felser2020efficient}, it is worth noting that the binary 2D TTN and the Hilbert curve mapping generate the same long range interactions within the corresponding networks. Therefore, the former benefits from the mathematical locality-preserving properties of the Hilbert curve. From this point of view, our analysis justifies 
the capability of binary 2D TTNs on the basis of the interaction range of 
the effective 1D model induced by the network structure. 

In conclusion, our work indicates a systematic strategy to evaluate, given a quantum many-body system defined on a two-dimensional lattice, the more efficient site-ordering to perform simulations by using TNs, based on the analysis of the effective, one-dimensional system generated by the mapping. In particular our work suggests that 1D variational ground state searching algorithms, such as DMRG, strongly benefits from the choice of the Hilbert curve for treating 2D systems. {\color{black} Even though in this work we focused on 2D square systems, our approach can be easily generalised to rectangular lattices, such as long narrow cylinders, since it is indeed possible to define the Hilbert curve also in this case}. Other possible outlooks include the application of this strategy to the study of systems characterized by non-trivial lattice structures and interactions, as for example in spin glasses \cite{Mydosh_2015} and quantum chemistry \cite{Cao2019}.

\section*{Acknowledgments}
We kindly acknowledge fruitful discussion with Timo Felser and Pietro Silvi. MPS calculations were performed using the TeNPy Library \cite{tenpy}.
This work is partially supported by the Italian PRIN2017 and Fondazione CARIPARO, the INFN project QUANTUM, the QuantERA projects QTFLAG and QuantHEP, the DFG project TWITTER, the Horizon 2020 research and innovation programme under grant agreement No 817482 (Quantum Flagship - PASQuanS). We 
acknowledge computational resources by the Cloud Veneto, CINECA, the BwUniCluster, ATOS Bull, and by University of Padova Strategic Research Infrastructure Grant 2017:  ``CAPRI: Calcolo ad Alte Prestazioni per la Ricerca e l'Innovazione''.
AA, NDP, and VG acknowledge support by MIUR via PRIN 2017 (Progetto di Ricerca di Interesse Nazionale): project QUSHIP (2017SRNBRK).

\appendix

\section{Choice of $\lambda=2.9$}\label{app:app_DE}

We have computed the energy difference $\Delta E$ for both the MPS and the TTN representatiosn with $m=20,35$ for values of $\lambda$ ranging in 
the interval $[0,4]$ which includes the critical point. Then we have chosen, as explained in the main text, the value of $\lambda$ where we observe a peak in the precision improvement. Note that a clear improvement emerges at low bond dimension values, implying that the Hilbert mapping guarantees a faster convergence of the ground state energy. The results are shown in Fig. \ref{fig:app_energy_gap}:
We have computed the energy difference $\Delta E$ at $n=8$ with OBC and 
PBC by using the TTN algorithm (first row) and at $n=16$ with OBC by using the MPS ground state representation (second row). In both rows, the left and right panels refer to the results obtained with $m=20,35$, respectively.

\section{Simulations with PBC}\label{app:PBC_TTN}

Here we show the results obtained by employing the TTN and the MPS Ansatze for the computation of the ground state with the Hilbert and the snake mappings. As anticipated in the main text, we observe an improvement when the Hilbert mapping is adopted, as shown in Figs. \ref{fig:app_PBC_E_gs_ttn} and \ref{fig:app_PBC_E_gs_mps}.

\section{Numerical simulation details}\label{app:numerical_details}
{\color{black}  All the numerical simulations have been performed with standard MPS-DMRG and Tree Tensor Networks variational ground-state searching algorithm. 

MPS-DMRG algorithm represent the state-of-the-art technique for the numerical simulation of many-body systems in 1D \cite{Schollw_ck_2011}. The calculations were performed by using the TeNPy Library \cite{tenpy} and by exploiting the parity-symmetry of the model. The error in the final energy at fixed bond dimension is below $10^{-6}$.

For the TTN algorithm we exploit the parity-simmetry of the Ising model and the Krylov sub space expansion \cite{TN_Anthology}. Given the model Hamiltonian, the TTN ground state is searched by iteratively optmizing each of the tensors in the network. By performing standard contraction operations, we derive a local eigenvalue problem for each tensor in the network, which is solved by using the Arnoldi method of the ARPACK library \cite{Lehoucq97arpackusers}. This procedure is iterated by sweeping through the TTN from the lowest to the highest layers, gradually reducing the energy expectation value. After completing the whole sweep, the procedure is iterated several times, until the desired convergence in the energy is reached. The convergence threshold of the final energy at fixed bond dimension is below $10^{-6}$.
}

\newpage 
\bibliographystyle{unsrtnat}
\bibliography{references}

\end{document}